\documentclass[journal]{IEEEtran}
\usepackage{amsmath,amssymb,amsthm}
\usepackage{graphicx}
\usepackage{cite}
\usepackage{bm}
\usepackage{algorithm}
\usepackage{algpseudocode}
\usepackage{subcaption}
\usepackage{flushend}
\usepackage{xcolor}
\newtheorem{theorem}{Theorem}

\begin{document}

\title{QIEN: Quantum-Inspired Energy-Neutral UAV Trajectory for WPT-Enabled Industrial HRLLC}

\author{Ali~Ranjha,
        Georges Kaddoum,~\IEEEmembership{Senior Member, IEEE},
         and Halim Yanikomeroglu~\IEEEmembership{Fellow, IEEE}
\thanks{
A. Ranjha and H. Yanikomeroglu are with the Carleton-NTN Lab at the the Department of Systems and Computer Engineering, Carleton University, Ottawa, Canada, e-mails: ali-nawaz.ranjha.1@ens.etsmtl.ca,  and halim@sce.carleton.ca, respectively.\\
G. Kaddoum is with the Department of Electrical Engineering, \'Ecole de Technologie Sup\'erieure, Montr\'eal, Canada, e-mail: Georges.Kaddoum@etsmtl.ca
}}  

\maketitle

\begin{abstract}
This paper introduces Quantum-Inspired Energy-Neutral (QIEN) UAV Trajectory for WPT-Enabled Industrial HRLLC, a novel framework for mobility-integrated trajectory and resource optimization in solar-powered uncrewed aerial vehicle (UAV) relay systems supporting autonomous mining robots in post-disaster environments. We formulate a mixed-integer nonlinear program (MINLP) that jointly optimizes UAV trajectory, finite blocklength allocation, transmit power, and wireless power transfer (WPT) over discrete time horizons to maximize aggregate data throughput. The system operates under strict extreme hyper-reliable low-latency communication (HRLLC) constraints, energy sustainability requirements via solar harvesting, and continuous robot power delivery. To solve this large-scale optimization problem, QIEN employs a hybrid Cooperative Quantum-behaved Particle Swarm Optimization (Co-QPSO) algorithm using quantum tunneling mechanisms and a heterogeneous swarm architecture. Our simulation results demonstrate that the proposed QIEN framework significantly outperforms benchmark algorithms, achieving successful convergence while maintaining energy-neutral operation and satisfying all reliability, power, and mobility constraints.
\end{abstract}


\section{Introduction}

\IEEEPARstart{T}{he} Industry 4.0 transformation of mining operations is fundamentally reshaping safety and efficiency management in extraction environments. Modern smart mining requires autonomous monitoring systems capable of operating in hazardous zones where human presence is unsafe due to toxic gas accumulation, unstable geological structures, and extreme temperatures \cite{liu2020industry,coutinho2022uav,onifade2023safe}. In these high-risk scenarios, autonomous ground robots are essential for perimeter surveillance, structural integrity assessment, and equipment inspection \cite{halder2023robots}. However, the rugged topography of open-pit mines frequently obstructs direct line-of-sight (LoS) communication, and post-disaster terrestrial networks are often compromised. To remedy this, Uncrewed Aerial Vehicles (UAVs) offer transformative solutions for reliable industrial data transmission \cite{xu2025trust,ranjan2020leveraging}. Acting as highly mobile aerial relays, UAVs bridge the connectivity gap by establishing flexible communication links that dynamically adapt to changing ground obstacles and robot positions \cite{zhang2017joint,zhan2011wireless,yao2025uav}.

To support real-time closed-loop control for catastrophic failure prevention, these UAV-assisted networks require extreme Hyper-Reliable Low-Latency Communication (HRLLC) \cite{kwak2025boosted}. Unlike conventional broadband services, mission-critical mining applications rely on the transmission of short control packets with stringent latency bounds \cite{tanveer2025spectral,tan2025caching}. Consequently, traditional network designs based on asymptotic Shannon capacity are inadequate. Instead, HRLLC necessitates finite blocklength (FBL) operation where transmission time is comparable to channel coherence time. In this regime, channel dispersion significantly impacts achievable rates, making the decoding error probability non-negligible. Therefore, the UAV relays must execute sophisticated resource allocation strategies to maintain stringent reliability guarantees—such as a $10^{-9}$ packet loss rate—without sacrificing spectral efficiency \cite{polyanskiy2010channel}.

\subsection{Related Work}
\label{sec:related_work}

However, using these adaptive UAV relays in such demanding environments presents several distinct technical challenges. To contextualize our proposed solution within this landscape, in this section, we review the literature across the following four critical domains: UAV channel modeling, energy-aware trajectory optimization, finite blocklength (FBL) communications, and metaheuristic optimization strategies.

\subsubsection{UAV Channel Modeling and Airborne Networks}
Accurate channel modeling is fundamentally essential for evaluating the performance of highly mobile UAV and aerial communication links. Recent comprehensive studies highlighted the distinct propagation characteristics of these environments. For instance, Cao \textit{et al.} \cite{cao2025survey} surveyed Near-Space Information Networks (NSINs), emphasizing the significant differences between High-Altitude Platform (HAP) and Autonomous Aerial Vehicle (AAV) channel models. The aforementioned work mathematically analyzed how the unstable movements of airborne platforms cause severe phase delays in onboard antenna arrays, impacting transmission reliability. Similarly, An \textit{et al.} \cite{puguang2025survey} investigated Airborne Maneuvering Networks (AMNs), highlighting that intrinsic channel characteristics--particularly, in High-Altitude Air-to-Ground (HA2G) and Air-to-Air (A2A) links--present unique propagation challenges that require specialized, robust modeling beyond traditional terrestrial assumptions. 

\subsubsection{Trajectory Optimization under Energy and WPT Constraints}
A significant body of recent literature has focused on overcoming the limited onboard energy of UAVs through trajectory optimization and Wireless Power Transfer (WPT) integration. In one relevant study, Wang \textit{et al.} \cite{wang2026drl} investigated multiple UAV base stations powered by wireless charging towers, using Deep Reinforcement Learning (DRL) to balance dynamic channel conditions and onboard energy limits for fair communication coverage. In the context of Simultaneous Wireless Information and Power Transfer (SWIPT), Baduge \textit{et al.} \cite{baduge2024flight} optimized UAV positioning in NOMA networks to maximize achievable rates while allowing the UAV to harvest energy from received signals. Furthermore, Tyrovolas \textit{et al.} \cite{tyrovolas2024energy} highlighted the distinct energy consumption characteristics of active UAV-mounted components, such as full-duplex relays (FDRs). The authors developed an energy-aware trajectory design that dynamically tightly couples propulsion power dynamics with the operational constraints of the relay to prevent mid-air energy depletion. 

\subsubsection{HRLLC and FBL Regimes}
HRLLC necessitates a departure from traditional Shannon capacity models, driving recent research to optimize UAV networks strictly within FBL regimes. For instance,
Bhattacharyya \textit{et al.} \cite{bhattacharyya2026q} developed an analytical framework for a UAV relay system employing Q-RSMA under an FBL regime, demonstrating that accounting for short-packet communications is essential to achieving $99.99\%$ reliability and sub-millisecond latency. Feng \textit{et al.} \cite{feng2026power} formulated a non-convex optimization problem for UAV-assisted photovoltaic plant inspections under URLLC constraints, minimizing transmission power through trajectory planning and bandwidth allocation despite imperfect channel state information (CSI). Additionally, Li \textit{et al.} \cite{li2026urllc} tackled URLLC constraints in Space-Air-Ground Integrated Networks (SAGINs) by utilizing IRS backscatter to reduce energy consumption, decomposing the problem using Lyapunov optimization and DRL.

\subsubsection{Algorithmic Positioning and Metaheuristic Optimization}
Solving the highly coupled mixed-integer nonlinear programming (MINLP) problems inherent in UAV networks frequently requires advanced optimization techniques. To date, classical metaheuristics have been widely applied in this domain: for instance, while Baduge \textit{et al.} \cite{baduge2024flight} deployed standard Particle Swarm Optimization (PSO) for UAV positioning, Qiao \textit{et al.} \cite{qiao2024resource} proposed a hybrid PSO-Genetic Algorithm (PSO-GA) to maximize fair sum rates in UAV-assisted high-speed rail systems under HRLLC requirements. 

While DRL \cite{wang2026drl, li2026urllc} and classical metaheuristics \cite{baduge2024flight, qiao2024resource} demonstrated their utility in specific UAV resource allocation problems, they present structural limitations when applied to discrete-continuous search spaces. DRL handles constraints via soft reward penalties, which cannot strictly guarantee the $10^{-9}$ HRLLC reliability required in our system. Meanwhile, traditional metaheuristics such as GA and standard PSO suffer from the ``invisible wall'' phenomenon; specifically, their Gaussian mutations and velocity-based Newtonian updates fail to bridge the severely disconnected feasible manifolds created by simultaneous FBL requirements, strict WPT delivery limits, and energy sustainability constraints. 

To overcome these Newtonian limitations, Sun \textit{et al.} \cite{sun2004particle, sun2004global} established the foundational concept of replacing particle velocity with a quantum delta potential well model. Building on this mathematical framework, quantum-inspired algorithms have recently emerged as powerful tools for UAV and mobile robot path planning. For instance, Huang \textit{et al.} \cite{huang2020novel} developed an improved QPSO using a particle dimension search strategy to accelerate global convergence in fixed-wing UAV route planning. Similarly, Liu \textit{et al.} \cite{liu2019uav} integrated QPSO with rolling horizon control (RHC-QPSO) to enable real-time dynamic obstacle avoidance in complex environments. More recently, Jing \textit{et al.} \cite{jing2025rl} proposed a hybrid RL-QPSO network, fusing Deep Reinforcement Learning with QPSO to bypass the local optima and high computational costs typically associated with deterministic search strategies in mobile robotics. 

However, despite these advancements, a critical limitation persists in the existing quantum-inspired UAV literature: namely, these implementations \cite{huang2020novel, liu2019uav, jing2025rl} inherently apply uniform quantum behavior or generalized learning policies across the entire population. When applied to the massive temporal coupling of multi-dimensional MINLPs (such as our HRLLC-WPT system), this uniform approach leads to rapid swarm diversity loss. The entire swarm frequently becomes trapped in local feasible “islands" bounded by vast infeasible spaces, causing premature stagnation.

By contrast, in this paper, we propose a Hybrid Cooperative Quantum-Behaved PSO (Co-QPSO). By abandoning velocity vectors in favor of a wavefunction-derived position update yielding a heavy-tailed Laplace spatial distribution, the core QPSO mechanics structurally guarantee a strictly higher mathematical probability of ``quantum tunneling'' through expansive infeasible regions \cite{sun2004particle}. Yet, the primary innovation of this work lies in the heterogeneous \textit{cooperative sub-swarm} architecture. Unlike the uniform swarms in \cite{huang2020novel, liu2019uav}, Co-QPSO explicitly divides the population to concurrently manage exploitation (via adaptive contraction coefficients) and aggressive exploration (via fixed hunter-particle mechanics). This architectural division resolves the issue of premature convergence flaws of standard quantum algorithms, thus significantly outperforming both classical architectures and baseline quantum methods in highly constrained, multi-objective trajectory planning.

\subsection{Novelty and Contributions}

To address the challenges outlined above, in this paper, we introduce \textbf{QIEN}, a framework for perpetual, reliable industrial monitoring. The main contributions are:

\begin{itemize}
    \item \textbf{Unified Energy-Neutral System Model:} We develop a holistic model coupling finite blocklength signaling with solar harvesting dynamics and WPT. In this regard, we explicitly model the ``Harvest-then-Transmit'' protocol under atmospheric attenuation and cloud shading, ensuring the trajectory is valid for both physical energy sustainability and cyber-physical reliability.

    \item \textbf{HRLLC-Constrained MINLP Formulation:} We formulate the joint trajectory and resource allocation problem as a large-scale Mixed-Integer Nonlinear Program (MINLP). The objective maximizes aggregate throughput while strictly enforcing HRLLC error probabilities, integer blocklength constraints, and a strict energy-neutral balance for the UAV operations.

    \item \textbf{Heterogeneous Cooperative Swarm Architecture:} {To solve the highly non-convex MINLP, in this study, we propose a novel cooperative swarm architecture that partitions computational agents into distinct ``Safety'' (exploitation) and ``Hunter'' (exploration) roles. To enable a smooth functioning of this architecture within the heavily constrained, discontinuous feasible region of our system model, we build it upon a heavy-tailed sampling foundation (Co-QPSO). This underlying stochastic sampling allows Hunter particles to bypass constraint barriers without the momentum-trapping failures characteristic of classical velocity-based methods.}

    \item \textbf{Demonstration of Perpetual Autonomy:} The results of our extensive simulations validate that QIEN outperforms classical benchmarks, achieving convergence while maintaining a sustainable energy margin and delivering continuous wireless power to support the ground robot during the mission.
\end{itemize}

\section{System Model and Problem Formulation}

\begin{figure}[!t]
\centering
\includegraphics[width=7cm, height=7cm]{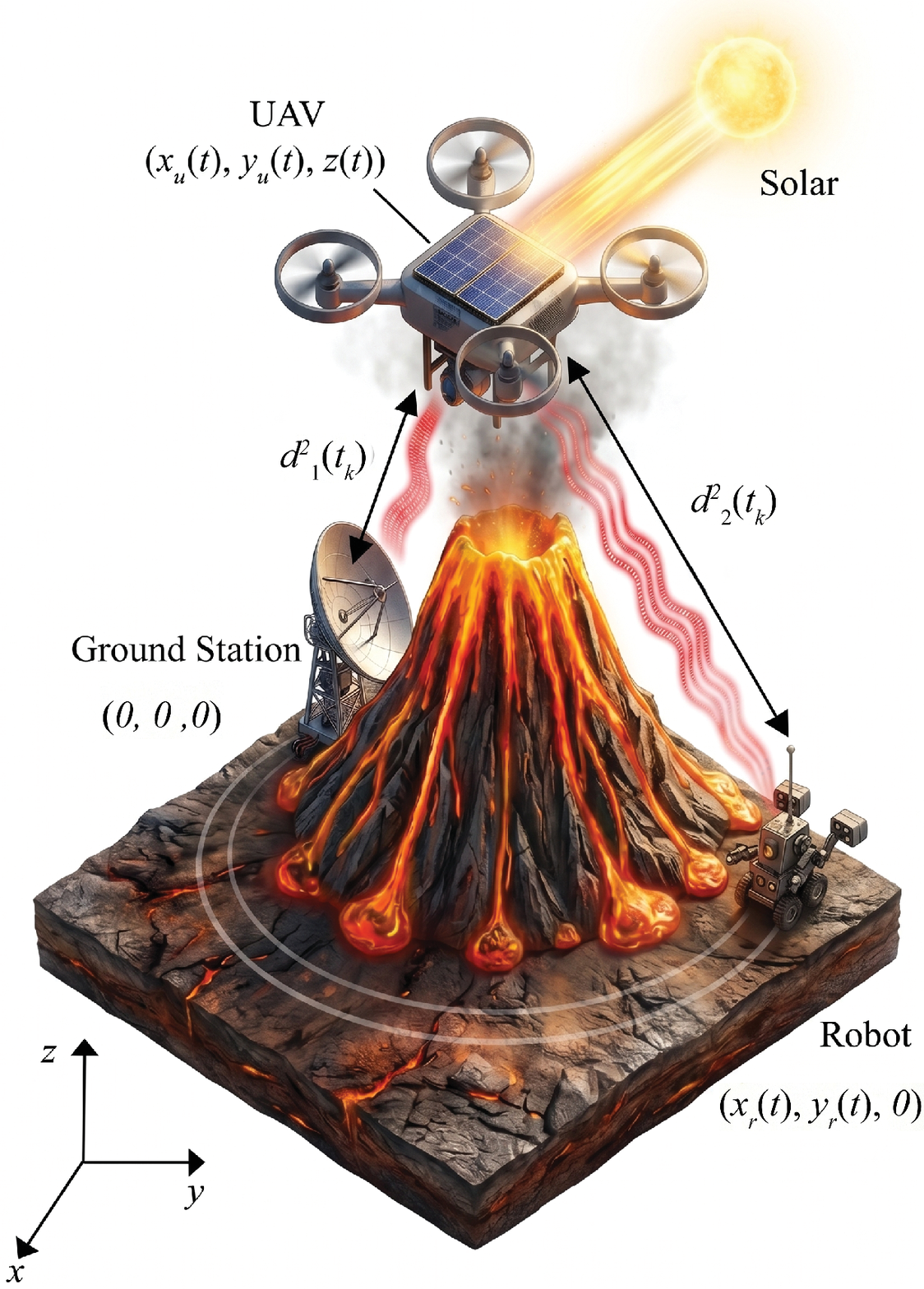}
\caption{QIEN: Solar-powered UAV relay and WPT architecture for HRLLC-driven robotics.}
\label{fig:system_model}
\end{figure}

We consider a wireless system comprising a ground station at fixed location $(0, 0, 0)$, a solar-powered UAV relay at time-varying position $(x_u(t), y_u(t), z(t))$, and a mobile inspection robot at ground position $(x_r(t), y_r(t), 0)$ following a prescribed circular trajectory. This architecture is designed to support the broad spectrum of HRLLC-driven robotics, enabling continuous autonomous operation in dynamic, infrastructure-denied, or hazardous environments. As a representative use case, we contextualize our model within Industry 4.0 smart mining operations, where autonomous robots perform continuous safety inspections in hazardous zones requiring HRLLC for real-time monitoring of mine boundaries, gas concentration mapping, and equipment health assessment. The UAV operates under a harvest-then-transmit protocol over mission duration $\mathcal{T} = [0, T_{\text{end}}]$, concurrently relaying information from the ground station to the robot while providing wireless power transfer to sustain robot operation.

Since, due to volcanic terrain and topography, the ground station and robot have no direct LoS, the UAV acts as an essential decode-and-forward (DF) relay for communication and concurrently delivers power. Importantly, the UAV-to-robot link does not employ a traditional Simultaneous Wireless Information and Power Transfer (SWIPT) architecture, which would necessitate complex power-splitting circuits at the receiver. Instead, we assume a dual-link, out-of-band architecture. The UAV uses a standard RF transceiver to forward HRLLC data blocks (using power $P_2$) over a dedicated communication band. Concurrently, the UAV is equipped with a highly directional energy transmitter that beams power ($P_{\text{tx}}^{\text{WPT}}$) to the ground robot over a distinct, lower-frequency WPT band (characterized by a higher channel gain $\beta_0^{\text{WPT}}$). Consequently, the ground robot is equipped with the following two independent, non-interfering hardware modules: an information decoding receiver and a dedicated rectenna array for pure energy harvesting. This architectural separation allows the system to independently optimize the communication and WPT transmit powers without suffering from the severe rate-energy trade-offs inherent to single-RF SWIPT receivers.

\subsection{Time Discretization and Mobility Model}

To make the continuous optimization problem tractable, we discretize the mission time $\mathcal{T}$ into $K$ uniform time steps $t_k = k \Delta t$ for $k \in \mathcal{K} = \{0, 1, \ldots, K-1\}$, where the total mission duration satisfies $T_{\text{end}} = (K-1)\Delta t$. All time-dependent system variables are evaluated at these discrete instances $\{t_k\}_{k \in \mathcal{K}}$. The time-varying communication distance for the ground-station-to-UAV link is given by the following:
\begin{equation}
d_1^2(t_k) = x_u^2(t_k) + y_u^2(t_k) + z^2(t_k),
\label{eq:d1}
\end{equation}
while the UAV-to-robot link distance depends on both UAV and robot positions:
\begin{equation}
d_2^2(t_k) = (x_u(t_k) - x_r(t_k))^2 + (y_u(t_k) - y_r(t_k))^2 + z^2(t_k).
\label{eq:d2}
\end{equation}
Assuming free-space propagation, the channel gains for both links is given as follows:
\begin{equation}
h_i(t_k) = \frac{\beta_0}{d_i^2(t_k)}, \quad i \in \{1, 2\}, \quad k \in \mathcal{K},
\label{eq:channel_gain}
\end{equation}
where $\beta_0$ represents the channel gain constant at unit reference distance. Here, the robot follows a predetermined circular trajectory for perimeter surveillance, with positions given by the following equations:
\begin{equation}
\begin{aligned}
x_r(t) &= x_c + r_c \cos(\theta_0 + \omega t), \\
y_r(t) &= y_c + r_c \sin(\theta_0 + \omega t),
\end{aligned}
\label{eq:robot_trajectory}
\end{equation}
where $(x_c, y_c)$ denotes the circle center, $r_c$ is the radius, $\omega$ is the angular velocity, and $\theta_0$ is the initial angular position. 

Furthermore, the UAV must remain within a maximum horizontal operational radius $X_{\max}$ and bounded flight altitudes $z \in [z_{\min}, z_{\max}]$. To ensure reliable LoS communication with the ground station, the UAV should maintain a minimum elevation angle $\theta_{\min}$:
\begin{equation}
z(t_k) \geq \sqrt{x_u^2(t_k) + y_u^2(t_k)} \tan(\theta_{\min}), \quad \forall k \in \mathcal{K}.
\label{eq:elevation_constraint}
\end{equation}

While recent advances in UAV channel modeling emphasize probabilistic LoS functions, such as elevation-dependent blockage models \cite{al2014optimal}, embedding these highly non-convex probability functions directly into the finite blocklength rate equations renders the massive MINLP mathematically intractable. To maintain tractability while respecting modern UAV channel behaviors, we adopt a bounded large-scale channel model governed by a strict geometric constraint (Eq. \ref{eq:elevation_constraint}). By dynamically forcing the UAV to maintain an elevation angle $\theta \geq \theta_{\min}$ above the ground nodes, the system mathematically guarantees operation within the dominant LoS regime characteristic of modern elevation-dependent empirical channel models, thereby isolating the algorithmic convergence performance without being obscured by stochastic channel variations.

The UAV's physical mobility is constrained by its maximum operational velocities. Since the proposed framework assumes a rotary-wing UAV capable of hovering for continuous inspection, its kinematic model exhibits decoupled aerodynamic limits for axial and translational flight. Specifically, vertical maneuvers must overcome gravitational forces yet are strictly bounded by total motor thrust, whereas horizontal cruising uses pitch dynamics to achieve significantly higher speeds. To accurately reflect these distinct mechanical bounds, the maximum horizontal and vertical velocities are independently constrained. Between consecutive time steps, the horizontal displacement must satisfy the following requirement:
\begin{equation}
\|(x_u(t_{k+1}), y_u(t_{k+1})) - (x_u(t_k), y_u(t_k))\|_2 \leq v_{\max}^{\text{horiz}} \Delta t,
\label{eq:horiz_mobility}
\end{equation}
for all $k \in \{0, \ldots, K-2\}$. Likewise, the vertical displacement is governed by a separate, stricter operational limit, represented by the constraint below:
\begin{equation}
|z(t_{k+1}) - z(t_k)| \leq v_{\max}^{\text{vert}} \Delta t, \quad \forall k \in \{0, \ldots, K-2\}.
\label{eq:vert_mobility}
\end{equation}
Consequently, the UAV's total 3D speed is implicitly bounded by the vector sum of these independent orthogonal constraints, which ensures that the generated trajectories remain physically executable for standard rotary-wing platforms. For operational safety, the UAV must also maintain a minimum separation distance $d_{\min}^{\text{GS}}$ from the ground station.

\subsection{Finite Blocklength Communication Model}
At each time step $t_k$, the UAV has a total of $N$ channel uses available for transmission. This blocklength must be allocated between the two communication links via an allocation vector $\bm{n}(t_k) = [n_1(t_k), n_2(t_k)]$, subject to the constraints $n_1(t_k) + n_2(t_k) = N$ and $n_i(t_k) \in \mathbb{Z}^+$. The transmit powers for the two links, denoted $P_1(t_k)$ and $P_2(t_k)$, are bounded by $[P_{\min}, P_{\max}]$.

Assuming an additive white Gaussian noise power of $\sigma^2 = N_0 B$, the instantaneous signal-to-noise ratio (SNR) for channel $i$ is $\gamma_i(t_k) = P_i(t_k) h_i(t_k) / \sigma^2$, yielding a Shannon capacity of $C_i(t_k) = \log_2(1 + \gamma_i(t_k))$. However, finite blocklength operation introduces stochastic variability quantified by channel dispersion:
\begin{equation}
V_i(t_k) = 1 - (1 + \gamma_i(t_k))^{-2}.
\label{eq:dispersion}
\end{equation}
According to Polyanskiy's finite blocklength theory \cite{polyanskiy2010channel}, the achievable rate in bits per second for link $i$ with target error probability $\varepsilon_{\text{target}}$ is approximated using the inverse Gaussian Q-function $Q^{-1}(\varepsilon) = \sqrt{2} \text{erfc}^{-1}(2\varepsilon)$:
\begin{equation}
R_i^*(t_k) = B\left[ C_i(t_k) - \log_2(e) \sqrt{\frac{V_i(t_k)}{n_i(t_k)}} Q^{-1}(\varepsilon_{\text{target}}) \right].
\label{eq:achievable_rate}
\end{equation}

To guarantee HRLLC, the total relay decoding error $\varepsilon_{\text{tot}}(t_k) = \varepsilon_1(t_k) + \varepsilon_2(t_k)$ must strictly satisfy $\varepsilon_{\text{tot}}(t_k) \leq \varepsilon_{\max}$ for all $k \in \mathcal{K}$, where we set $\varepsilon_{\max} = 10^{-9}$. Integrating the instantaneous sum rate $R_{\text{sum}}(t_k) = R_1^*(t_k) + R_2^*(t_k)$ over the entire mission duration, the aggregate data throughput is:
\begin{equation}
\bar{R}_{\text{sum}} = \Delta t \sum_{k=0}^{K-1} R_{\text{sum}}(t_k) = \Delta t \sum_{k=0}^{K-1} \sum_{i=1}^{2} R_i^*(t_k).
\label{eq:agg_throughput}
\end{equation}

\subsection{Energy Harvesting and Power Consumption Model}

The solar-powered UAV harvests energy that varies with altitude \cite{lee2017optimal}. The instantaneous harvested power at altitude $z(t_k)$ is modeled as shown below:
\begin{equation}
P_{\text{harvest}}(t_k) = \max\{A_{\text{solar}} \phi(z(t_k)) C(z(t_k)), 0\},
\label{eq:harvest_power}
\end{equation}
where the solar aperture constant $A_{\text{solar}} = \eta_{\text{solar}} S_p G$ combines panel efficiency, irradiance, and gain. The elevation angle factor accounts for atmospheric attenuation that decreases with altitude:
\begin{equation}
\phi(z) = \alpha - \beta_s e^{-z/\delta},
\label{eq:elevation_factor}
\end{equation}
where $\alpha$ is the base elevation factor, $\beta_s$ is the atmospheric decay coefficient, and $\delta$ is the characteristic atmospheric length scale. Moreover, the cloud layers introduce additional attenuation modeled through a piecewise function as shown below:
\begin{equation}
C(z) = \begin{cases}
e^{-\beta_c (L_{\text{up}} - L_{\text{low}})}, & z \leq L_{\text{low}}, \\
e^{-\beta_c (L_{\text{up}} - z)}, & L_{\text{low}} < z \leq L_{\text{up}}, \\
1, & z > L_{\text{up}},
\end{cases}
\label{eq:cloud_attenuation}
\end{equation}
where $L_{\text{low}}$ and $L_{\text{up}}$ define the cloud layer boundaries and $\beta_c$ is the cloud attenuation coefficient. 

On the consumption side, UAV propulsion power increases with altitude due to varying air density:
\begin{equation}
P_{\text{prop}}(t_k) = P_0 + P_1 z(t_k) + P_2 z^2(t_k),
\label{eq:propulsion}
\end{equation}
where $P_0$, $P_1$, and $P_2$ are propulsion coefficients. The total UAV power consumption $P_{\text{consume}}(t_k)$ aggregates propulsion, communication transmission powers ($P_1, P_2$), WPT transmission $P_{\text{tx}}^{\text{WPT}}$, and a constant circuit processing power $P_{\text{circuit}}$. {To ensure practical robustness, our energy causality model includes $\rho_{es}$ which denotes an Energy Safety Margin. Instead of forcing power consumption to exactly equal harvested power, the constraint now strictly reserves a percentage of the harvested energy to act as a realistic buffer, as shown below:
\small
\begin{align}
P_{\text{consume}}(t_k) &= P_{\text{prop}}(t_k) + P_1(t_k) + P_2(t_k) + P_{\text{tx}}^{\text{WPT}}(t_k) + P_{\text{circuit}} \notag \\
&\leq (1 - \rho_{\text{es}}) P_{\text{harvest}}(t_k), \quad \forall k \in \mathcal{K}.
\label{eq:consumption_margin}
\end{align}\normalsize}

\subsection{Wireless Power Transfer Model}
The UAV transmits power to the robot via radio frequency with a time-varying transmit power bounded by $P_{\text{tx}}^{\text{WPT}}(t_k) \in [0, P_{\text{tx},\max}^{\text{WPT}}]$. The power received at the robot depends on the transmit power, WPT-specific channel gain $\beta_0^{\text{WPT}}$, and RF-to-DC conversion efficiency $\eta_{\text{WPT}} \in (0,1)$:
\begin{equation}
P_{\text{WPT}}(t_k) = \frac{\eta_{\text{WPT}} P_{\text{tx}}^{\text{WPT}}(t_k) \beta_0^{\text{WPT}}}{d_2^2(t_k)}.
\label{eq:wpt_received}
\end{equation}
Due to lower operating frequencies, specialized rectenna arrays, and beam focusing techniques, WPT systems typically achieve higher effective channel gains than data communication links. To ensure continuous robot operation, the received power must satisfy $P_{\text{robot}} \leq P_{\text{WPT}}(t_k) \leq P_{\text{WPT}}^{\max}$, where the base requirement $P_{\text{robot}} = P_{\text{robot,circuit}} + P_{\text{robot,move}}$ combines circuit and locomotion components, and the upper bound prevents damage to the robot's energy harvesting circuitry.

\subsection{Problem Formulation}

With all system components defined, we now formulate the mobility-integrated trajectory optimization problem. Our objective is to maximize the aggregate data throughput over the entire mission:
\begin{subequations}\small
\begin{align}
\text{P1:} \quad \max_{\bm{\Phi}} \quad & \bar{R}_{\text{sum}}, \label{eq:obj}\\
\text{s.t.} \quad 
& n_1(t_k) + n_2(t_k) = N, \label{eq:c1}\\
& n_i(t_k) \in \mathbb{Z}^+, \quad i \in \{1,2\}, \quad \forall k \in \mathcal{K}, \label{eq:c2}\\
& x_u^2(t_k) + y_u^2(t_k) \leq X_{\max}^2, \quad \forall k \in \mathcal{K}, \label{eq:c3}\\
& z_{\min} \leq z(t_k) \leq z_{\max}, \quad \forall k \in \mathcal{K}, \label{eq:c4}\\
& z(t_k) \geq \sqrt{x_u^2(t_k) + y_u^2(t_k)} \tan(\theta_{\min}), \quad \forall k \in \mathcal{K}, \label{eq:c5}\\
& \varepsilon_{\text{tot}}(t_k) \leq \varepsilon_{\max}, \quad \forall k \in \mathcal{K}, \label{eq:c6}\\
& P_{\text{consume}}(t_k) \leq (1 - \rho_{\text{es}}) P_{\text{harvest}}(t_k), \quad \forall k \in \mathcal{K}, \label{eq:c7}\\
& P_{\text{WPT}}(t_k) \geq P_{\text{robot}}, \quad \forall k \in \mathcal{K}, \label{eq:c8}\\
& P_{\text{WPT}}(t_k) \leq P_{\text{WPT}}^{\max}, \quad \forall k \in \mathcal{K}, \label{eq:c9}\\
& P_{\min} \leq P_i(t_k) \leq P_{\max}, \quad i \in \{1, 2\}, \quad \forall k \in \mathcal{K}, \label{eq:c10}\\
& 0 \leq P_{\text{tx}}^{\text{WPT}}(t_k) \leq P_{\text{tx},\max}^{\text{WPT}}, \quad \forall k \in \mathcal{K}, \label{eq:c11}\\
& \|(x_u(t_{k+1}), y_u(t_{k+1})) - (x_u(t_k), y_u(t_k))\|_2 \notag \\
& \qquad \leq v_{\max}^{\text{horiz}} \Delta t, \quad \forall k \in \{0, \ldots, K-2\}, \label{eq:c12}\\
& |z(t_{k+1}) - z(t_k)| \leq v_{\max}^{\text{vert}} \Delta t, \notag \\
& \qquad \forall k \in \{0, \ldots, K-2\}, \label{eq:c13}\\
& d_1(t_k) \geq d_{\min}^{\text{GS}}, \quad \forall k \in \mathcal{K}, \label{eq:c14}
\end{align}
\end{subequations}\normalsize
where the optimization variable set encompasses all time-varying decision variables:
\begin{equation}
\small
\bm{\Phi} = \{x_u(t_k), y_u(t_k), z(t_k), n_1(t_k), P_1(t_k), P_2(t_k), P_{\text{tx}}^{\text{WPT}}(t_k)\}_{k \in \mathcal{K}}.
\end{equation}

Problem P1 is a large-scale MINLP with $9K$ decision variables, comprising $7K$ continuous variables and $2K$ integer variables (blocklength allocations). The problem exhibits severe non-convexity that arises from multiple sources: integer constraints \eqref{eq:c2}, nonlinear time-varying channel gains in \eqref{eq:obj}, the inverse Q-function in the finite blocklength rate formula, bilinear terms in the WPT constraints \eqref{eq:c8}-\eqref{eq:c9}, and temporal coupling through mobility constraints \eqref{eq:c12}-\eqref{eq:c13} (see Appendix A). Classical gradient-based methods fail due to integer constraints and non-convex objective structure, while global optimization methods suffer from the curse of dimensionality with an increase of $K$ as well as the ``invisible wall'' phenomenon (See Appendix B). This motivates our quantum-inspired metaheuristic approach.

\section{Hybrid Cooperative Quantum-Behaved PSO}

\subsection{Classical PSO Limitations}

Classical Particle Swarm Optimization (PSO) has been widely applied to continuous optimization problems \cite{clerc2002particle}. In classical PSO, each particle $j$ maintains position $\mathbf{\Phi}_j$ and velocity $\mathbf{v}_j$, updated according to the following:
\begin{equation}
\mathbf{v}_j(i+1) = w\mathbf{v}_j(i) + c_1 r_1 (\mathbf{\Phi}_j^{\text{pb}} - \mathbf{\Phi}_j) + c_2 r_2 (\mathbf{\Phi}^{\text{gb}} - \mathbf{\Phi}_j),
\end{equation}
\begin{equation}
\mathbf{\Phi}_j(i+1) = \mathbf{\Phi}_j(i) + \mathbf{v}_j(i+1),
\end{equation}
where $w$ is inertia weight, $c_1$ and $c_2$ are cognitive and social coefficients, respectively, $r_1, r_2 \sim U(0,1)$ are random parameters, and $\mathbf{\Phi}_j^{\text{pb}}$ and $\mathbf{\Phi}^{\text{gb}}$ denote personal and global best positions, respectively.

However, when applied to Problem P1, classical PSO exhibits a fundamental limitation known as the invisible wall phenomenon. This occurs in heavily constrained search spaces where multiple constraints interact nonlinearly. Specifically, when energy sustainability constraint \eqref{eq:c7}, WPT requirements \eqref{eq:c8}-\eqref{eq:c9}, and mobility limits \eqref{eq:c12}-\eqref{eq:c13} are concurrently active, the feasible region becomes a narrow, disconnected manifold embedded in high-dimensional space. Classical PSO particles accumulate momentum through the velocity term, causing overshooting behavior that repeatedly violates multiple constraints. Once trapped in infeasible regions, the velocity-based update mechanism reinforces movement patterns that lead to constraint violations, rather than facilitating escape. The gradient information from penalty functions provides insufficient guidance to navigate back to feasibility.

\subsection{Quantum-Behaved PSO Foundation}

To overcome these limitations, Quantum-behaved PSO (QPSO) fundamentally redesigns particle dynamics by entirely eliminating the velocity term. Instead, QPSO models particles as quantum entities with probabilistic position distributions, inspired by quantum mechanics where particles exist in superposition states \cite{mikki2006quantum}. In QPSO, particles update positions through a local attractor mechanism combined with quantum uncertainty. For each particle $j$ and dimension $d$ at iteration $i$, the local attractor position is computed as a random convex combination of the particle's personal best and the global best:
\begin{equation}
p_j^{(d)}(i) = \phi_j^{(d)} \Phi_j^{\text{pb},(d)} + (1 - \phi_j^{(d)}) \Phi^{\text{gb},(d)},
\end{equation}
where $\phi_j^{(d)} \sim U(0,1)$ randomly balances the influence of personal and global best positions. The particle position is then updated according to the following:
\begin{equation}
\Phi_j^{(d)}(i+1) = p_j^{(d)}(i) + s \alpha(i) \left| m_{\text{best}}^{(d)}(i) - \Phi_j^{(d)}(i) \right| \ln\left(\frac{1}{u_j^{(d)}}\right),
\end{equation}
where $s \in \{-1, +1\}$ is randomly chosen with equal probability, $u_j^{(d)} \sim U(0,1)$ introduces stochasticity, and $\alpha(i)$ is the contraction-expansion coefficient. The mean best position guides global exploration as shown below:
\begin{equation}
m_{\text{best}}^{(d)}(i) = \frac{1}{N_p} \sum_{j=1}^{N_p} \Phi_j^{\text{pb},(d)}.
\end{equation}

Logarithmic term $\ln(1/u_j^{(d)})$ models quantum tunneling behavior, producing occasional large jumps that enable particles to traverse infeasible barriers without accumulating problematic momentum. This mechanism mimics quantum particles penetrating energy barriers that are classically forbidden, thus allowing escape from local optima and navigation through complex constraint landscapes \cite{grifoni1998driven,isakov2016understanding,le2011quantum}. 

It is important to clarify that while QPSO and our proposed Co-QPSO are categorized as ``quantum-inspired'' algorithms, they are fundamentally classical metaheuristics that do not require quantum computing hardware to execute. The terminology originates strictly from the mathematical foundation: rather than utilizing the Newtonian kinematics of traditional PSO, the state of a particle is depicted by a wavefunction derived from the Schr\"{o}dinger equation in a Delta potential well. Computationally, this manifests as the heavy-tailed (Laplace) probability distribution used to sample new particle positions. This heavy-tailed stochastic sampling is the actual mechanism driving the extensive global exploration and ``quantum tunneling'' phenomena \cite{sun2004particle}.

As such, the contraction-expansion coefficient linearly decreases over iterations to balance exploration and exploitation:
\begin{equation}
\alpha(i) = 1.0 - 0.5 \frac{i}{I_{\max}}.
\label{eq:contraction}
\end{equation}

\subsection{Cooperative Swarm Architecture}

While standard QPSO applies uniform quantum behavior across all particles, this can lead to either excessive exploration preventing convergence or premature exploitation missing global optima. Our Hybrid Cooperative QPSO (Co-QPSO) addresses this concern through heterogeneous particle roles that concurrently explore and exploit different regions of the search space.

The swarm of $N_p$ particles is partitioned into two equal-sized cooperative sub-swarms with distinct behaviors. Safety particles, comprising the first half of the swarm, prioritize exploitation by using the adaptive contraction coefficient $\alpha_{\text{safe}}(i)$ from (\ref{eq:contraction}). In addition, their local attractor weight undergoes a smoothing transformation as shown below:
\begin{equation}
\phi_j^{(d)}(i) = 0.5 + (r_j^{(d)} - 0.5) \left(1 - \frac{i}{I_{\max}}\right), \quad r_j^{(d)} \sim U(0,1),
\end{equation}
where the factor $(1 - i/I_{\max})$ gradually reduces randomness as iterations progress. This causes $\phi_j$ to converge towards 0.5, creating symmetric balance between personal and global best influences and promoting stable convergence in final iterations.

Hunter particles, comprising the second half, maintain aggressive exploration throughout optimization by using a fixed contraction coefficient $\alpha_{\text{hunter}} = 0.7$ regardless of iteration count. Their attractor weights remain fully random with $\phi_j^{(d)} \sim U(0,1)$, thus enabling continuous probing of distant regions even after safety particles have converged. This heterogeneous architecture prevents premature stagnation while allowing simultaneous refinement of promising solutions.

When the global best solution enters the feasible region (fitness $< 10^7$), Co-QPSO applies perturbation-based local search to refine the solution. At each iteration, two dimensions are randomly selected and perturbed according to the following:
\begin{equation}
\Phi^{\text{gb},(d)} \leftarrow \Phi^{\text{gb},(d)} + \mathcal{N}(0,1) \rho(\Phi_{\max}^{(d)} - \Phi_{\min}^{(d)}) \left(1 - \frac{i}{I_{\max}}\right),
\end{equation}
where $\rho = 0.05$ is the perturbation ratio and $\mathcal{N}(0,1)$ is Gaussian noise. The perturbation magnitude decreases with iterations, implementing adaptive precision that transitions from coarse adjustments to fine-grained refinement. Limiting perturbations to two dimensions prevents destructive changes while allowing targeted improvements.

\begin{figure}
    \centering
    \includegraphics[width=1\linewidth]{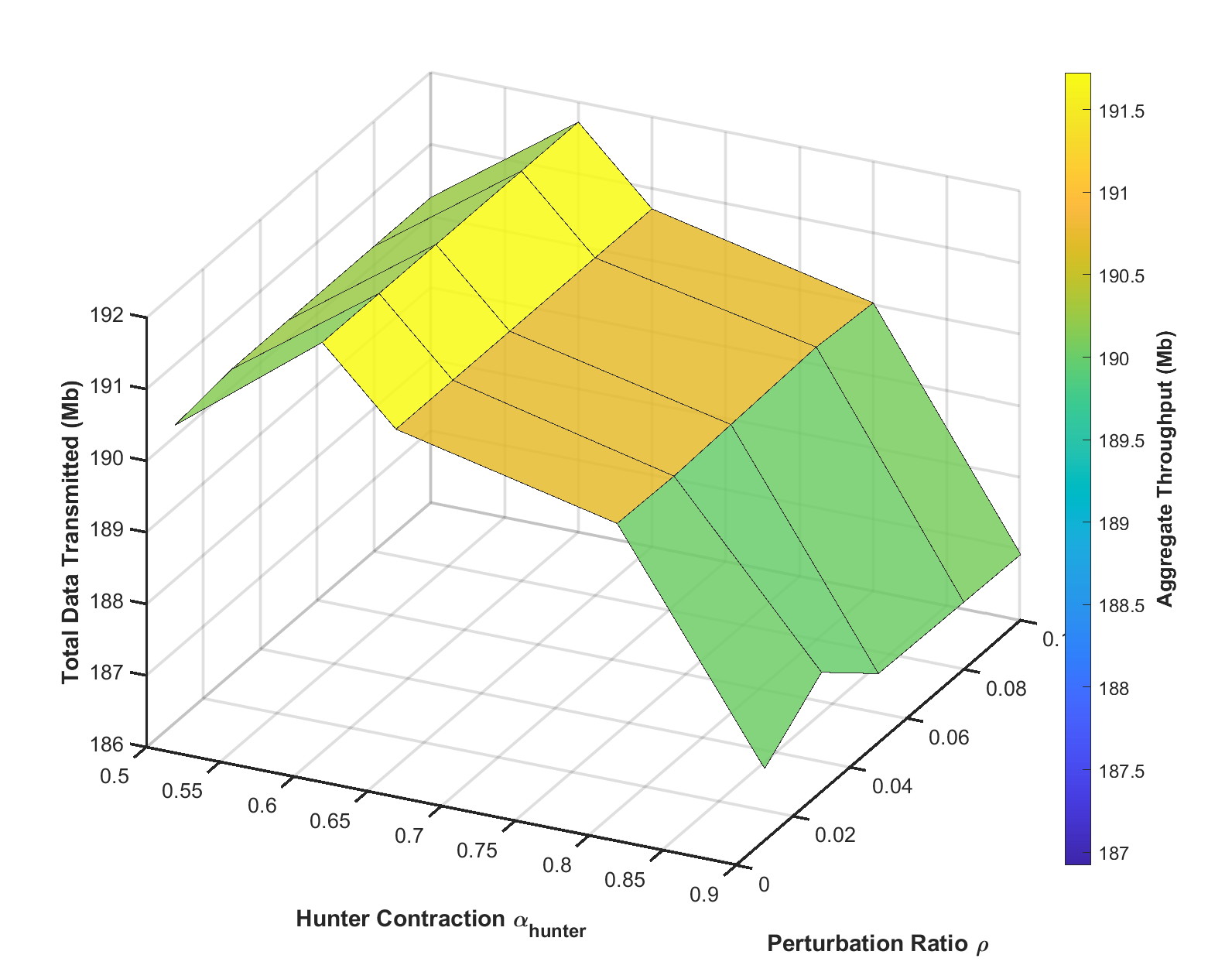}
    \caption{Sensitivity analysis.}
    \label{fig:sensitivity}
\end{figure}

\subsection{Fitness Function, Constraint Handling, and Quantum Terminology}

The fitness function transforms the maximization problem into minimization while handling constraints through adaptive penalties. For feasible solutions satisfying all constraints (\eqref{eq:c1}-\eqref{eq:c14}), the fitness equals the negative aggregate sum rate. For infeasible solutions, a large penalty base of $10^7$ is added to accumulated constraint violations, thus creating clear separation between feasible and infeasible regions. This penalty structure is implemented in the objective function evaluation within Algorithm 1, incorporating spatial boundary violations, energy sustainability deficits, WPT requirement shortfalls, mobility constraint violations, and ground station safety distance breaches. The specific penalty formulations are detailed in the implementation, with weights scaled according to constraint criticality: namely, energy sustainability and robot power delivery receive highest weights ($10^5$) due to system survival requirements, while spatial and mobility violations use moderate weights ($10^3$-$10^4$) to guide search without overwhelming objective improvement.

Importantly, while Co-QPSO is categorized as a `quantum-inspired' or `quantum-behaved' algorithm, it is fundamentally a classical metaheuristic. It does not require quantum computing hardware to execute. The term originates from its mathematical foundation: instead of using the Newtonian kinematics (velocity and trajectory) of traditional PSO, Co-QPSO models particle states using wavefunctions derived from the Schr\"{o}dinger equation in a Delta potential well. Computationally, this manifests itself as a heavy-tailed Laplace probability distribution for sampling new particle positions. This heavy-tailed sampling mechanism allows the algorithm to perform extensive global exploration and escape local optima—a stochastic process referred to in the optimization literature as `quantum tunneling' \cite{sun2004particle}.

\begin{algorithm}[t]
\caption{Hybrid Cooperative QPSO for Problem P1}
\label{alg:coqpso}
\begin{algorithmic}[1]
\State \textbf{Input:} $N_p, I_{\max}, \alpha_h, \rho$
\State \textbf{Init:} $\bm{\Phi}_j \sim \mathcal{U}(\bm{\Phi}_{\min}, \bm{\Phi}_{\max})$; Round integers; Find $\bm{\Phi}^{\text{gb}}$
\For{$i = 1$ to $I_{\max}$}
    \State $\mathbf{m}_{\text{best}} \gets \frac{1}{N_p}\sum \bm{\Phi}_j^{\text{pb}}$; \quad $\alpha_s \gets 1 - 0.5(i/I_{\max})$
    \For{$j = 1$ to $N_p$}
        \If{$j \le N_p/2$} \Comment{Safety Particle}
            \State $\alpha \gets \alpha_s$; $\phi \sim \mathcal{U}(0.5, 0.5 \pm \text{decay})$
        \Else \Comment{Hunter Particle}
            \State $\alpha \gets \alpha_h$; $\phi \sim \mathcal{U}(0,1)$
        \EndIf
        \For{$d = 1$ to $D$}
            \State $p \gets \phi \Phi_j^{\text{pb},(d)} + (1-\phi) \Phi^{\text{gb},(d)}$
            \State $\Phi_j^{(d)} \gets p \pm \alpha |m_{\text{best}}^{(d)} - \Phi_j^{(d)}| \ln(1/u)$
        \EndFor
        \State Clip bounds; Round integers; Update $\bm{\Phi}_j^{\text{pb}}$ if better
    \EndFor
    \State Update $\bm{\Phi}^{\text{gb}}$; \textbf{If} Feasible \textbf{then} Perturb $\bm{\Phi}^{\text{gb}}$
\EndFor
\State \textbf{Return} $\bm{\Phi}^{\text{gb}}$
\end{algorithmic}
\end{algorithm}

\section{Simulation Results}
\begin{figure}
    \centering
    \includegraphics[width=1\linewidth]{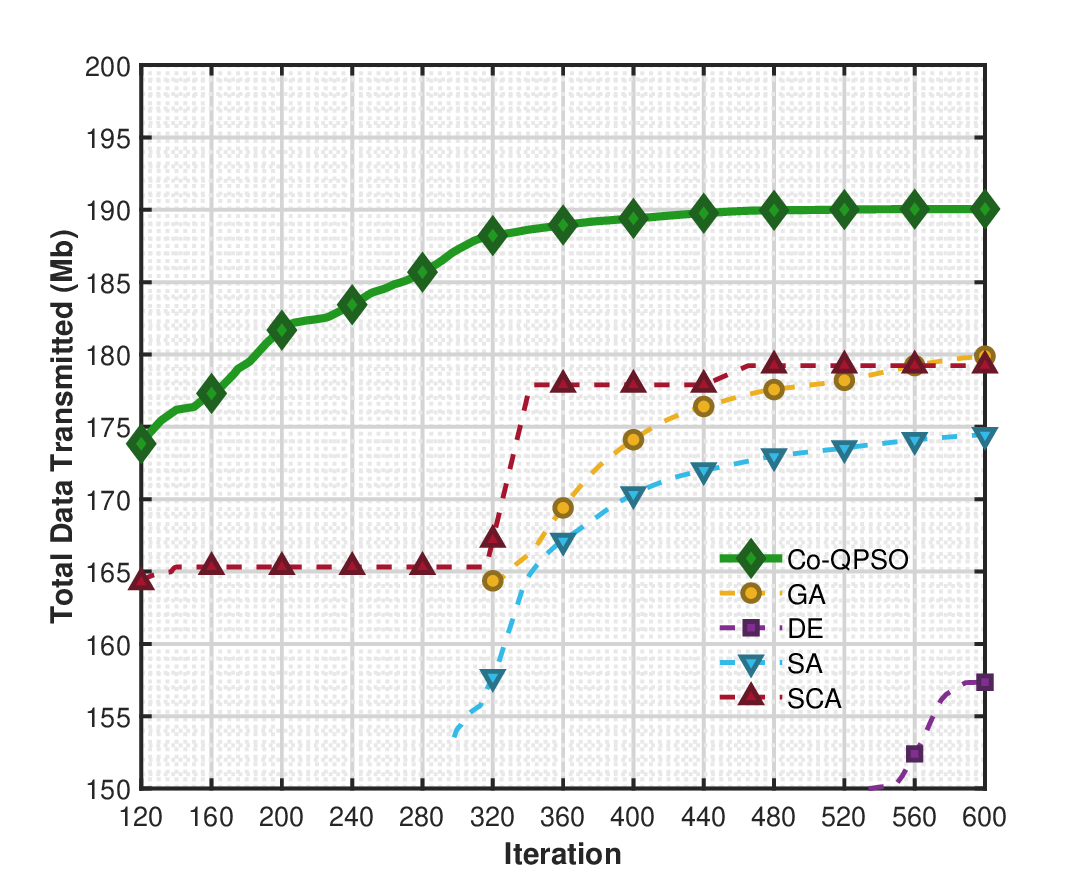}
    \caption{Benchmark comparisons.}
    \label{fig:benchmark}
\end{figure}

\begin{figure*}[!t]
\centering
\subfloat[Convergence comparison.]{\includegraphics[width=0.48\textwidth]{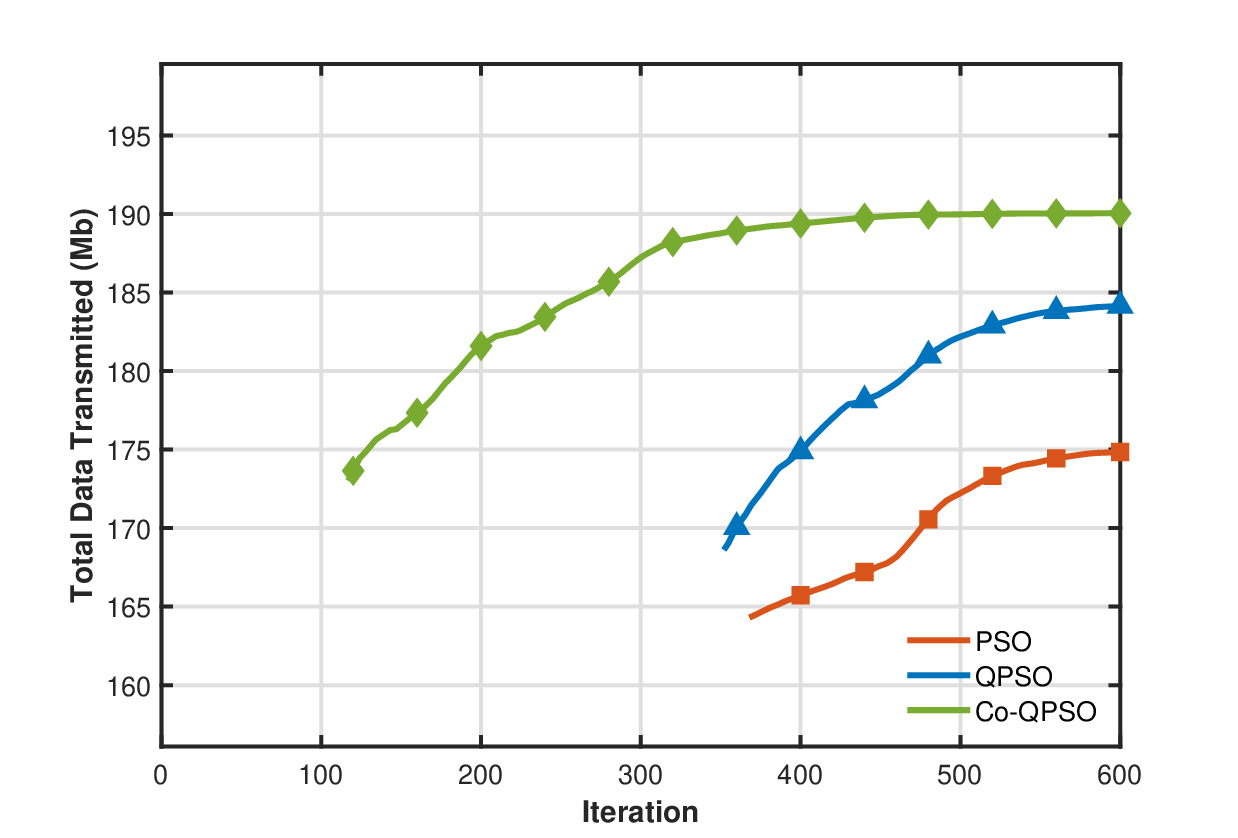}\label{fig:conv}}
\hfill
\subfloat[3D trajectory visualization.]{\includegraphics[width=0.48\textwidth]{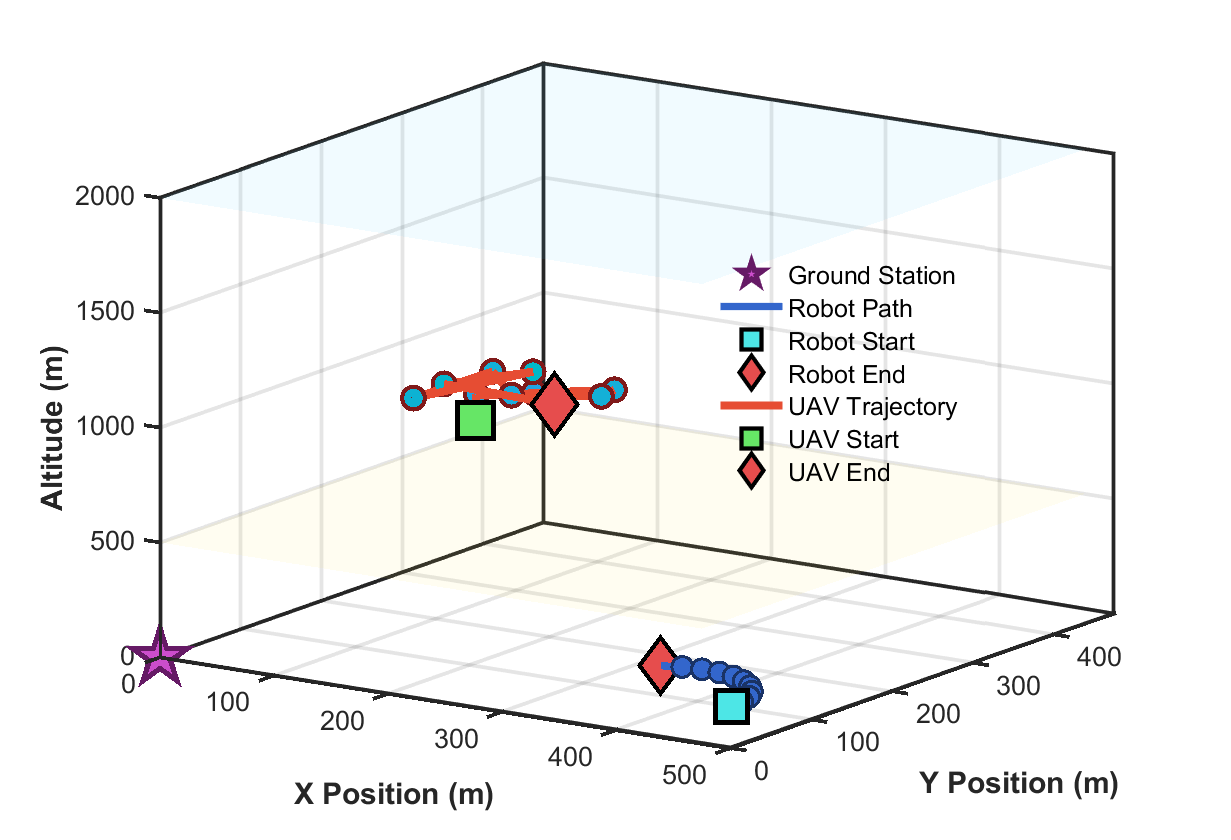}\label{fig:3d}}
\\
\subfloat[2D trajectory projection.]{\includegraphics[width=0.5\textwidth]{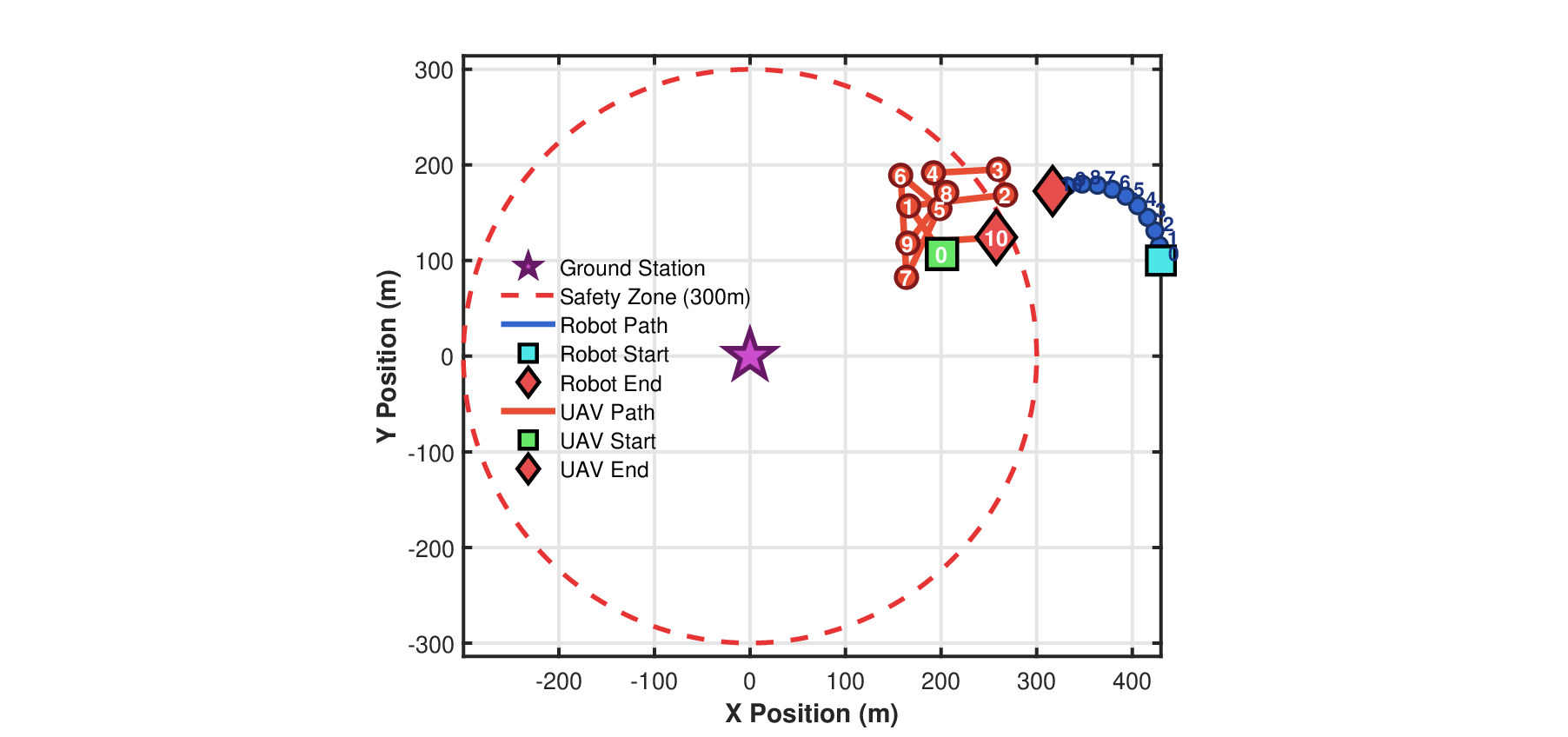}\label{fig:2d}}
\hfill
\subfloat[Solar harvesting vs WPT efficiency.]{\includegraphics[width=0.48\textwidth]{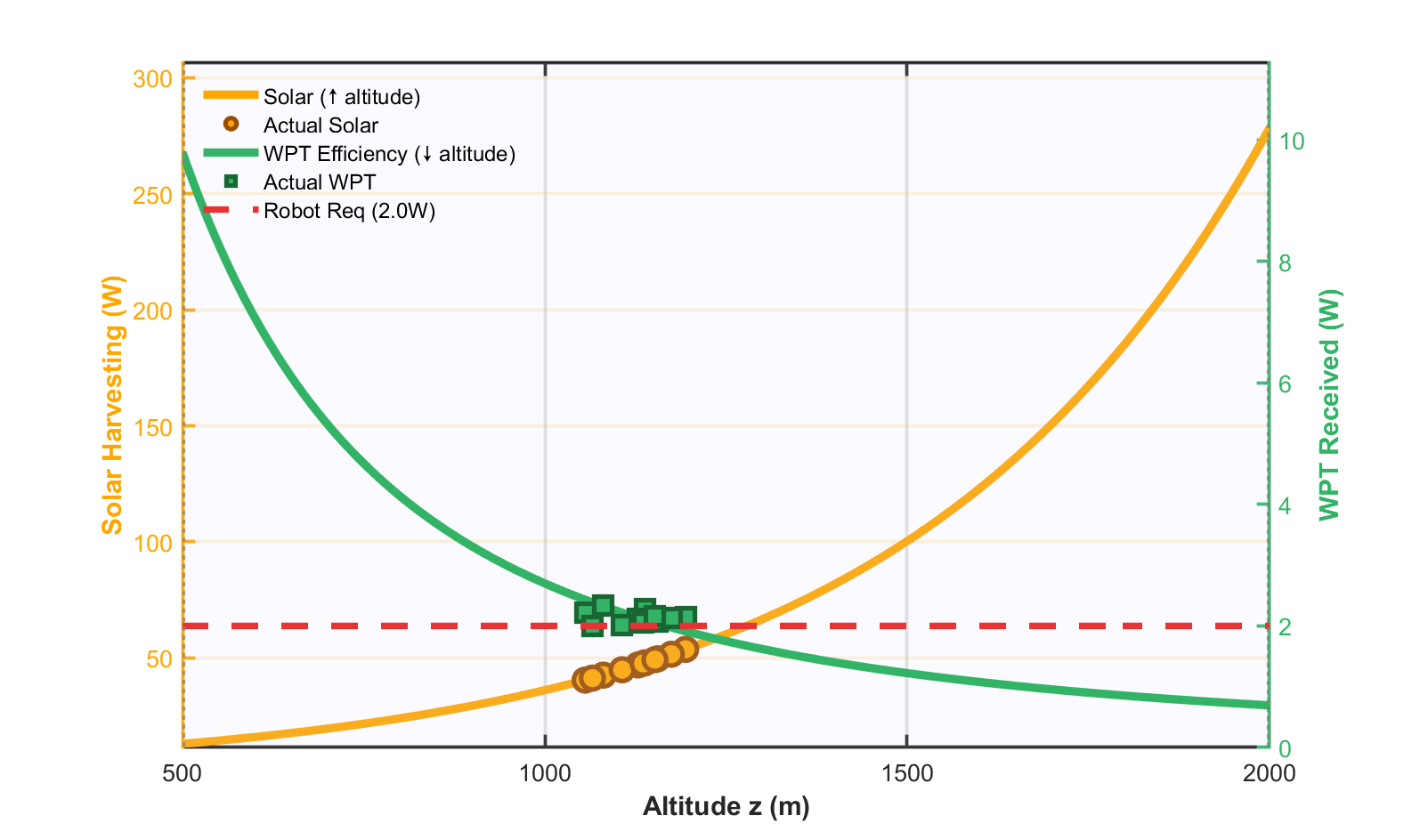}\label{fig:SolarvsWPT}}
\\
\subfloat[Energy-neutral balance.]{\includegraphics[width=0.48\textwidth]{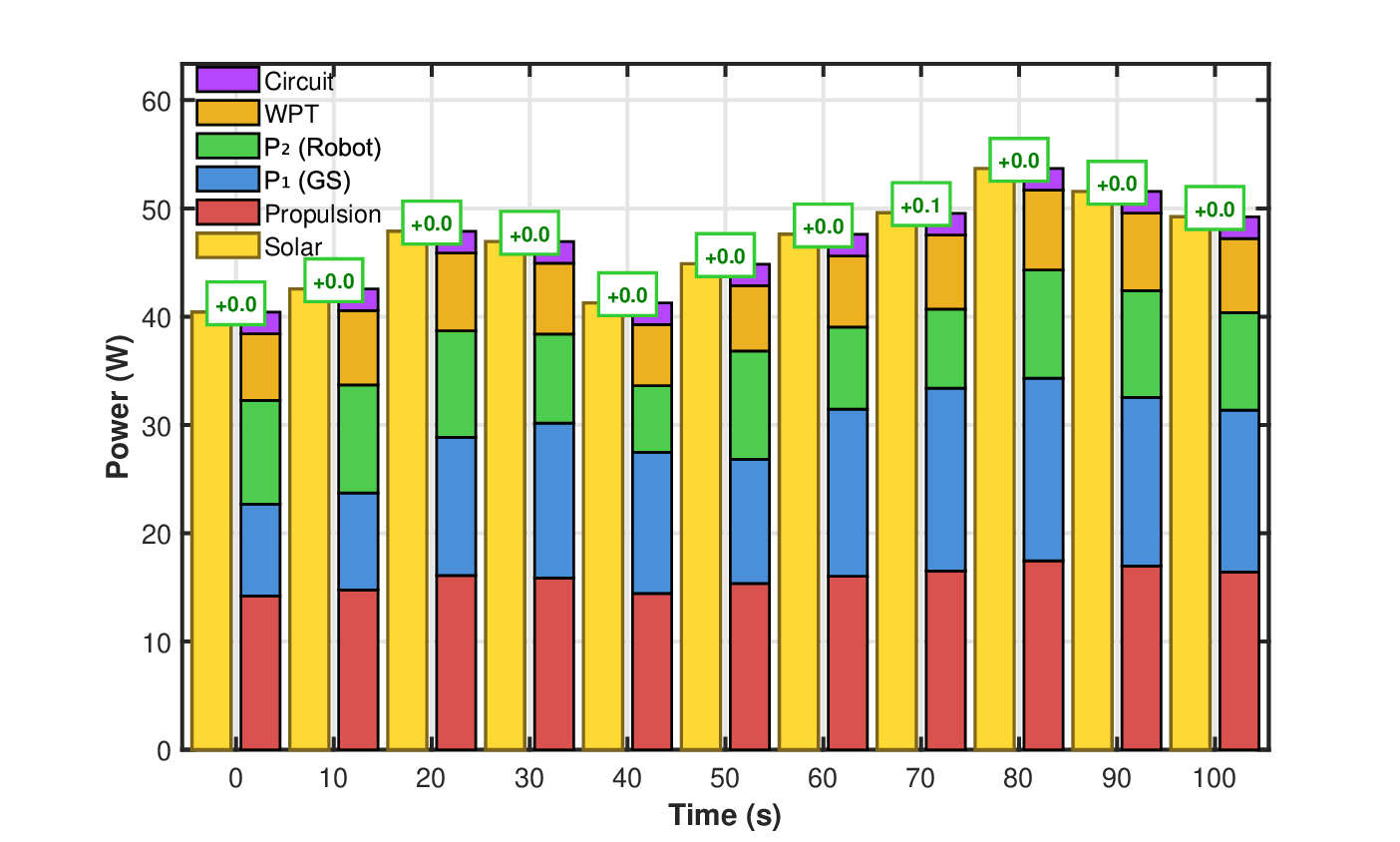}\label{fig:energy}}
\hfill
\subfloat[WPT performance over time.]{\includegraphics[width=0.48\textwidth]{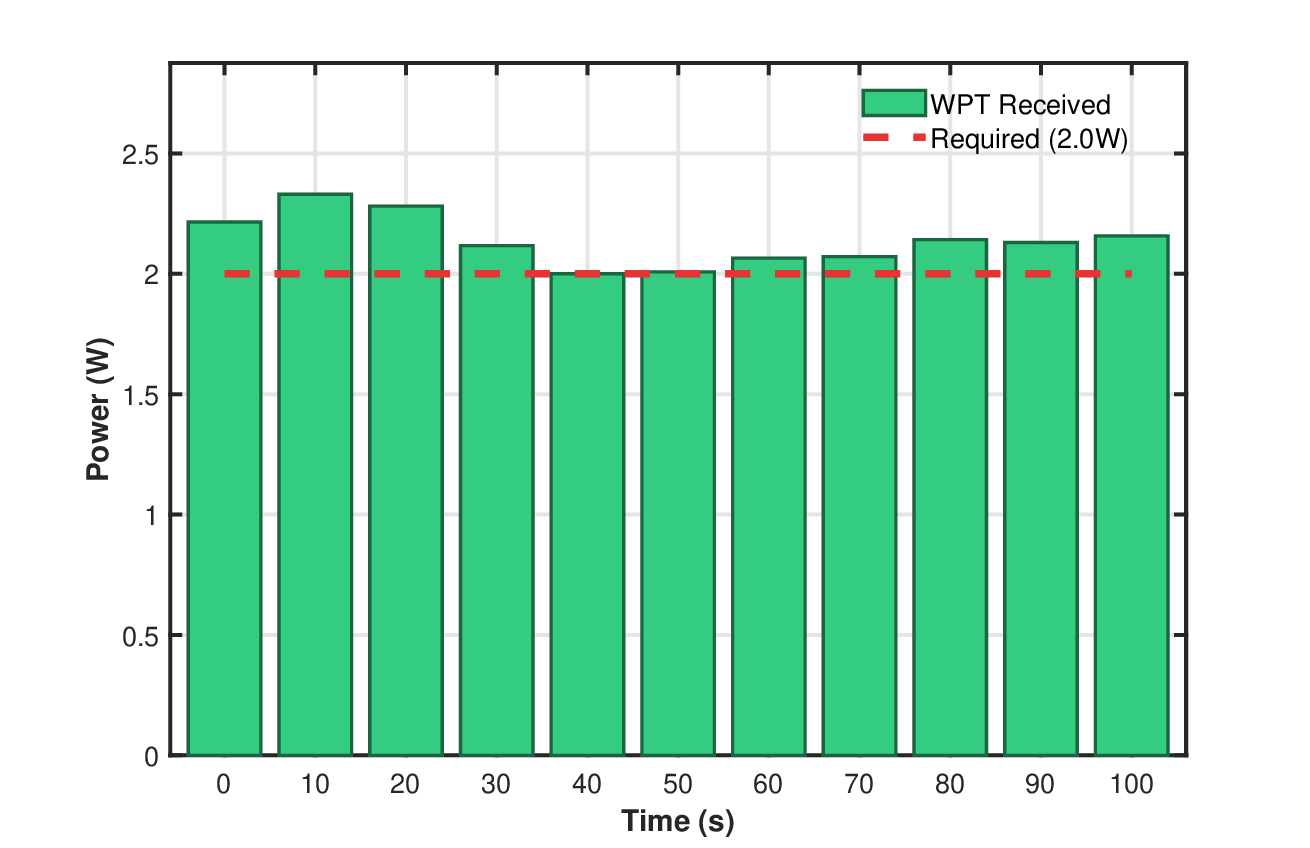}\label{fig:wpt}}
\caption{Performance evaluation of Co-QPSO algorithm: (a) convergence behavior; (b)-(c) trajectory optimization results; (d) trade-off analysis between solar harvesting and WPT efficiency; and (e)-(f) energy management analysis.}
\label{fig:results}
\end{figure*}

We evaluated the proposed Co-QPSO algorithm for a 100-second autonomous surveillance mission. The mission was discretized into $K = 11$ time steps with interval $\Delta t = 10$ seconds, resulting in 77 optimization variables. The robot executed a circular perimeter surveillance trajectory with center at $(350, 100)$ meters, radius $r_c = 80$ meters, and angular velocity $\omega = 0.02$ rad/s. System parameters were configured as follows: 
UAV altitude bounds $z \in [500, 2000]$ meters, maximum horizontal velocity $v_{\max}^{\text{horiz}} = 20$ m/s, maximum vertical velocity $v_{\max}^{\text{vert}} = 8$ m/s, transmit power bounds $P_1 \in [1, 22]$ W and $P_2 \in [1, 10]$ W, WPT transmit power bound $P_{\text{tx}}^{\text{WPT}} \in [0, 10]$ W, robot power requirement $P_{\text{robot}} = 2.0$ W, maximum WPT received power $P_{\text{WPT}}^{\max} = 2.5$ W, and HRLLC error constraint $\varepsilon_{\max} = 10^{-9}$. The Co-QPSO algorithm was configured with swarm size $N_p = 100$ particles and maximum iterations $I_{\max} = 600$. We evaluated performance with total blocklength $N = 150$ channel uses per time step.

\subsection{Hyperparameters and Sensitivity Analysis}

{The linearly decreasing contraction-expansion coefficient $\alpha(i)$ decaying from 1.0 to 0.5 applied to the Safety particles was directly adopted from foundational QPSO global convergence theory, previously introduced as the `creativity coefficient' $\beta$ by Sun \textit{et al.} \cite{sun2004global}. This decay mathematically guarantees that the particle state eventually collapses to the local attractor. However, the heterogeneous roles and custom parameters are novel to our framework. The Hunter particle coefficient ($\alpha_{\text{hunter}} = 0.7$) was empirically selected as the optimal equilibrium point between the boundaries of divergence ($\alpha > 1.0$) and rapid exploitation ($\alpha \to 0.5$). To prevent destructive tunneling that would displace the global best solution out of the narrow feasible manifolds, the perturbation ratio ($\rho = 0.08$) was maintained to be deliberately small (8\% of the domain).

To empirically prove robustness of the Co-QPSO framework, we conducted a sensitivity analysis on the following two critical custom hyperparameters: the Hunter contraction coefficient ($\alpha_{\text{hunter}}$) and the perturbation ratio ($\rho$). To eliminate deterministic bias and capture true algorithmic behavior, every point in the $5 \times 5$ grid search ($\alpha_{\text{hunter}} \in [0.5, 0.9]$ and $\rho \in [0.01, 0.10]$) was evaluated as the average of independent Monte Carlo trials using randomized seeds.

As shown in Fig. \ref{fig:sensitivity}, the algorithm exhibited remarkable stability. Every parameter combination successfully navigated the severe penalty landscape to find a feasible trajectory. The absolute maximum variance in the average final throughput across the entire grid amounted to only 2.51\%, ranging from 186.92 Mb to 191.72 Mb. This negligible variance convincingly demonstrates that the framework relies on its structural cooperative architecture to solve the MINLP, rather than brittle parameter tuning. 

Furthermore, surface topology explicitly revealed operational hierarchy of the parameters. The primary cooperative search ($\alpha_{\text{hunter}}$) dictates global convergence. Conversely, the perturbation ratio ($\rho$) acts strictly as an emergency fail-safe; while remaining inactive and yielding flat ridges when optimal $\alpha_{\text{hunter}}$ values efficiently prevent stagnation, but it actively influences the trajectory when suboptimal $\alpha$ values force the swarm into local minima.}

\subsection{Baseline Comparisons and Complexity Analysis}

{To accurately evaluate the proposed Co-QPSO framework, we benchmarked its convergence and throughput against the following four distinct optimization algorithms: Successive Convex Approximation (SCA) representing classical gradient-based methods, Genetic Algorithm (GA), Simulated Annealing (SA), and Differential Evolution (DE) representing classical stochastic metaheuristics. Since analytical gradients were unavailable, the SCA baseline is implemented as a derivative-free proximal-linear variant utilizing finite-difference gradients and an Armijo line search \cite{7882695}. To ensure a fair comparison within the heavily constrained search space, the evolutionary baselines were augmented with Deb's constraint-dominance rules. All algorithms were allocated an identical evaluation budget of 600 iterations with a population size of 100.

Fig. \ref{fig:benchmark} presents convergence graphs, revealing the following three distinct behavioral phenomena that validate the proposed approach:

\subsubsection{Gradient Trapping in SCA} 
While SCA used gradient information to rapidly ascend the penalty landscape by finding an early feasible solution by iteration 200, it ultimately fell victim to the severe non-convexity of the MINLP. As shown by the convergence curves, SCA exhibited a stair-step pattern that permanently flatlined at 179.24~Mb. Because of the lack of a stochastic mechanism for broad exploration, the gradient-based approach became irreversibly trapped in a local optimum.

\subsubsection{Delayed Feasibility in Standard Metaheuristics} 
Conversely, classical evolutionary algorithms (GA, SA, and DE) avoid permanent stagnation, yet suffer from severely delayed feasibility. GA and SA exhausted two-thirds of their computational budget (400 iterations) navigating infeasible regions, which illustrates the ``invisible wall'' phenomenon. Because their standard transition probabilities rapidly decay (e.g., Gaussian mutation), GA and SA struggled to bridge the disconnected feasible manifolds created by the tightly coupled WPT and energy constraints. DE experienced the most extreme delay, failing to locate a valid trajectory until the final evaluation window (iteration 600), which yielded a highly suboptimal throughput of 157.36~Mb.

\subsubsection{Advantages of the Proposed Co-QPSO}
The proposed Co-QPSO architecture resolved limitations of both baseline families. Driven by heavy-tailed Laplace sampling, the swarm penetrated the highly constrained feasible region as rapidly as SCA. Furthermore, the heterogeneous cooperative design prevented premature stagnation. The persistent global exploration of the ``Hunter'' particles, paired with the local exploitation of the ``Safety'' particles, enabled continuous trajectory refinement. Ultimately, Co-QPSO converged to a maximum aggregate throughput of 190.06~Mb, thus strictly satisfying all HRLLC and energy-neutrality requirements; by contrast, classical baselines yielded a 5.3\% to 17.2\% lower sum-rate as compared to the proposed architecture.}

Computational complexity of the evaluated optimization algorithms is primarily governed by the problem dimensionality $D = 7K$, population size $N_p$, maximum number of iterations $T_{\max}$, and cost of evaluating the objective function. For the time-integrated objective function, computing the trajectory-dependent channel capacities, energy consumption, WPT constraints, and temporal mobility penalties requires iterating over $K$ discrete time steps, yielding an evaluation complexity of $\mathcal{O}(K)$. For the population-based metaheuristics—including the proposed Co-QPSO, Genetic Algorithm (GA), Differential Evolution (DE), and the parallel-chain Simulated Annealing (SA)—each of the $N_p$ individuals undergoes a position update or reproductive operation taking $\mathcal{O}(D) = \mathcal{O}(K)$ time, which is followed by a single objective function evaluation per iteration. Consequently, the worst-case time complexity for Co-QPSO, GA, DE, and SA is strictly bounded by $\mathcal{O}(N_p \cdot T_{\max} \cdot K)$. By contrast, the SCA algorithm, implemented here via projected gradient descent, relies on finite differences to approximate the gradient over all $D$ dimensions. This requires $D$ objective function evaluations per iteration. Since $D \propto K$ and each evaluation costs $\mathcal{O}(K)$, the gradient computation scales quadratically, resulting in a worst-case time complexity of $\mathcal{O}(T_{\max} \cdot K^2)$ for SCA. As the mission duration or discretization density increases, this quadratic dependence on the number of time steps makes the gradient-based SCA significantly more computationally burdensome than metaheuristic approaches.

\vspace{-1.5mm}
\subsection{Optimality Analysis}

{To perform optimality analysis, we established a \textbf{Three-Tier Upper Bound Hierarchy} by strategically relaxing constraints as described below:

\begin{enumerate}
    \item \textbf{UB-Shannon (Theoretical Maximum):} We relaxed the finite blocklength (FBL) constraint by assuming infinite blocklengths, which is Shannon capacity limit, and removed the UAV mobility constraints. This allowed the solver to optimally place the UAV at each time slot independently, representing an absolute ceiling.
    \item \textbf{UB-FBL (Realistic Physical Limit):} While restoring the strict HRLLC FBL rate equations, we allowed continuous blocklength allocation and maintained the relaxed UAV mobility constraints. This provided a much tighter, physically realistic upper bound against to judge our algorithm against.
    \item \textbf{Proposed Co-QPSO (Achieved):} Our actual solution strictly enforced all integer blocklength constraints and physical UAV trajectory/velocity limits over the entire mission time.
\end{enumerate}
\begin{figure}
    \centering
    \includegraphics[width=1\linewidth]{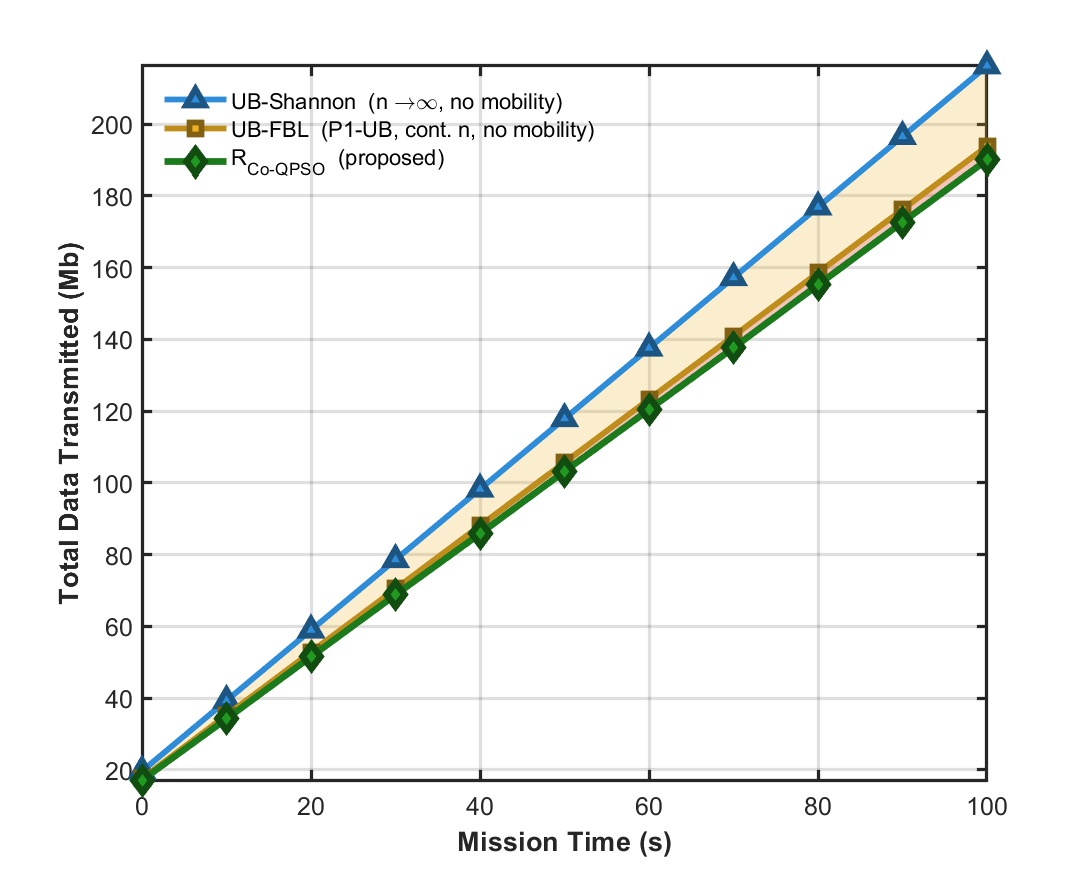}
    \caption{Optimality analysis vs upper bounds}
    \label{fig:optimality_1}
\end{figure}

\begin{figure}
    \centering
    \includegraphics[width=1\linewidth]{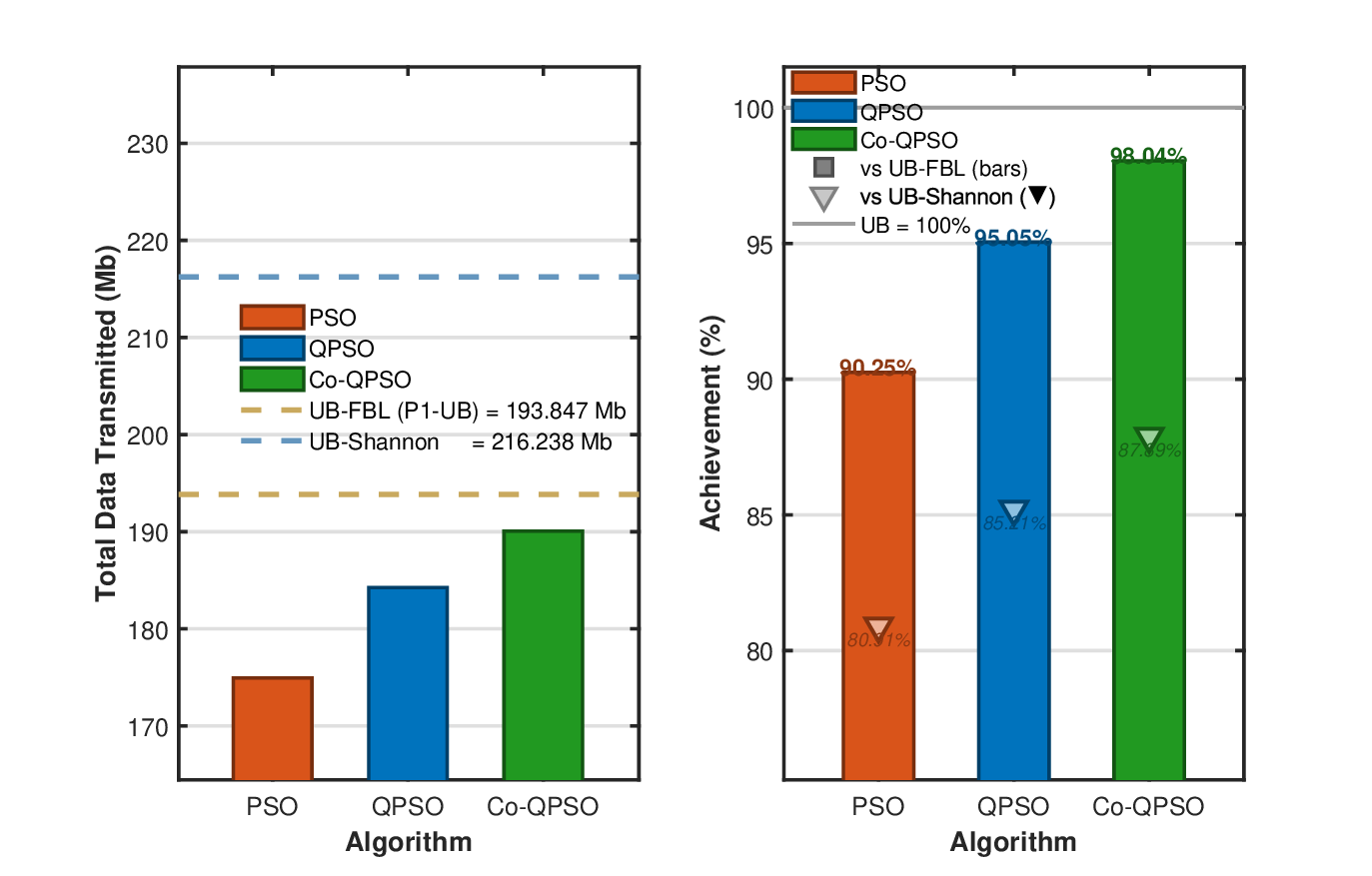}
    \caption{Optimality analysis vs upper bounds bar graph}
    \label{fig:optimality_2}
\end{figure}

Next, instead of presenting a single, monolithic optimality gap, we decomposed the total performance loss into two distinct components to clearly isolate algorithmic efficiency from fundamental physical limitations. To this end, the FBL overhead ($\delta_{\text{FBL}}$) first quantifies the gap between the UB-Shannon and UB-FBL bounds. Accounting for an unavoidable $10.35\%$ capacity reduction, this represents an inherent mathematical penalty of transmitting short packets under strict HRLLC error constraints—a fundamental physical layer characteristic, rather than an algorithmic inefficiency. Second, the algorithmic optimality gap ($\delta_{\text{Opt}}$) measures the difference between the realistic UB-FBL limit and our achieved Co-QPSO solution. This captures the actual performance cost of our optimization heuristics, integer blocklength restrictions, and the physical reality of continuous UAV flight. Remarkably, as shown in Fig. \ref{fig:optimality_1}-\ref{fig:optimality_2}, this algorithmic gap was merely $1.96\%$, demonstrating that the proposed Co-QPSO solution successfully achieves $98.04\%$ of the tightest theoretical bound (UB-FBL). Despite these strict physical layer constraints, the proposed solution still captured $87.89\%$ of the idealized, infinite-blocklength Shannon capacity limit.}
\vspace{1.5mm}
\subsection{Robustness Analysis}

\begin{figure}
    \centering
    \includegraphics[width=0.9\linewidth]{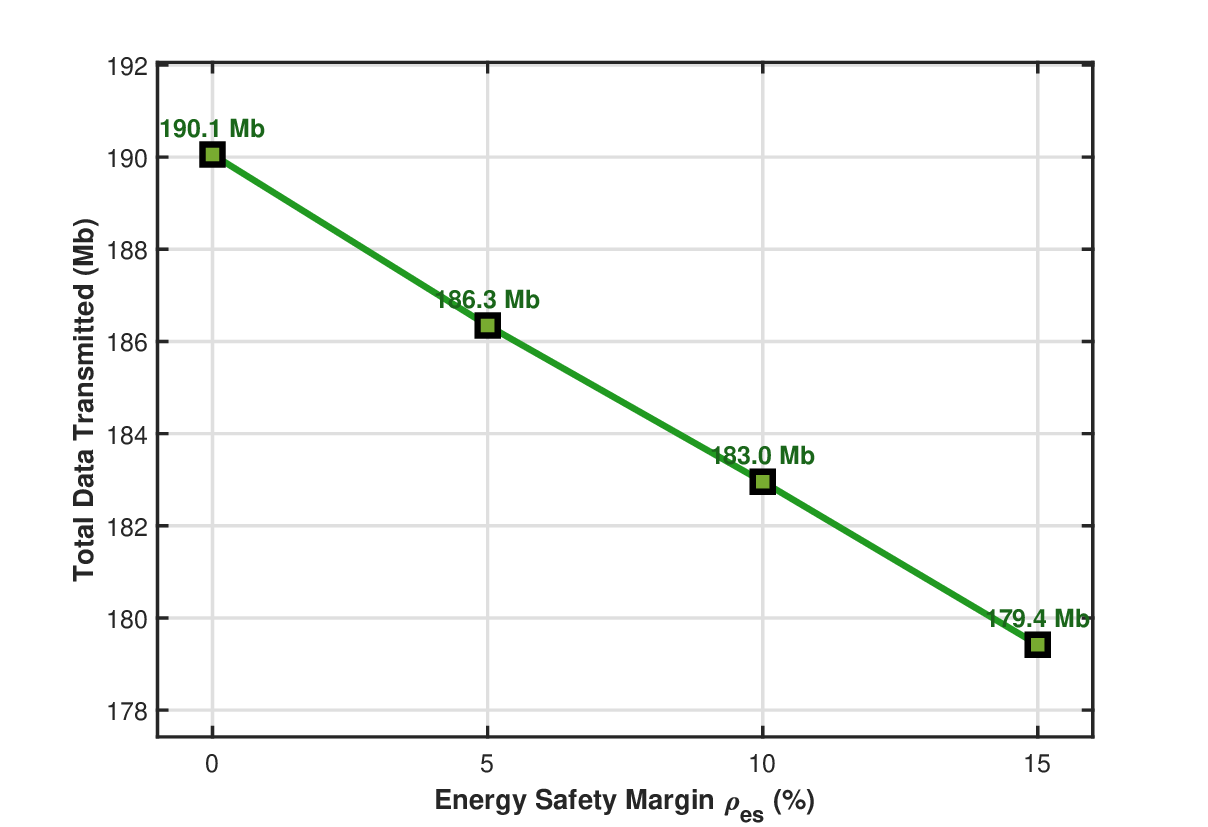}
    \caption{Robustness analysis for different energy safety margins}
    \label{fig:robustness}
\end{figure}

{To explicitly evaluate the robustness of our proposed Co-QPSO algorithm under these conditions, we added a new robustness analysis. We also varied the safety margin $\rho_{es}$ from $0\%$, which is the original idealized baseline to $15\%$, which translates to a highly conservative real-world buffer. As shown in Fig. \ref{fig:robustness}, we added a trade-off curve showing that the algorithm dynamically adapted to the restricted energy budget and gracefully degrades. The total transmitted data shifted from $190.06$ Mb (at $0\%$ margin) to $186.35$ Mb (at $5\%$), $182.96$ Mb (at $10\%$), and, ultimately, $179.42$ Mb at a strict $15\%$ margin. This represents merely a $5.60\%$ reduction in maximum throughput.} 

\begin{figure}[h]
    \centering
    \includegraphics[width=1\linewidth]{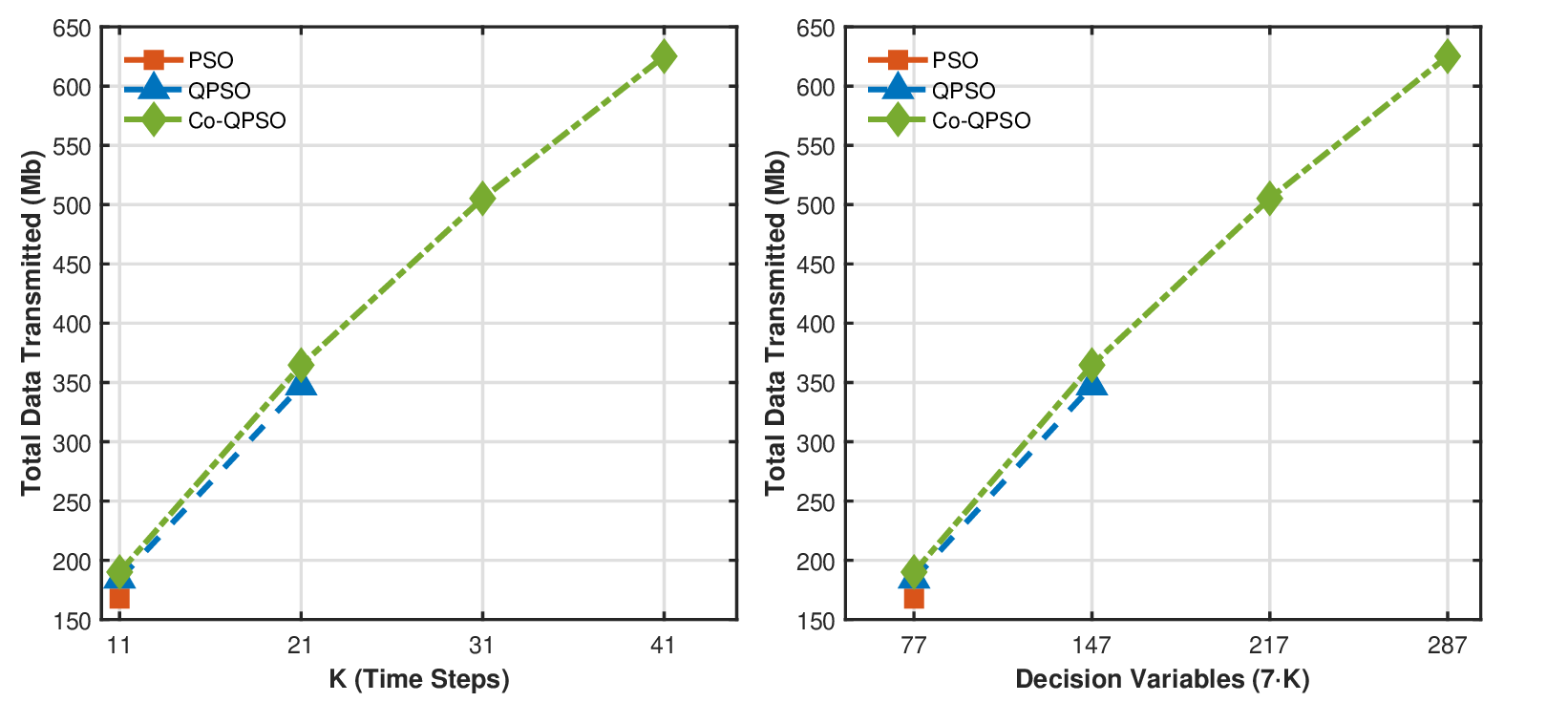}
    \caption{Scalability of PSO, QPSO, and Co-QPSO with extended $K$}
    \label{fig:Scalability_1}
\end{figure}

\begin{figure}[h]
    \centering
    \includegraphics[width=0.75\linewidth]{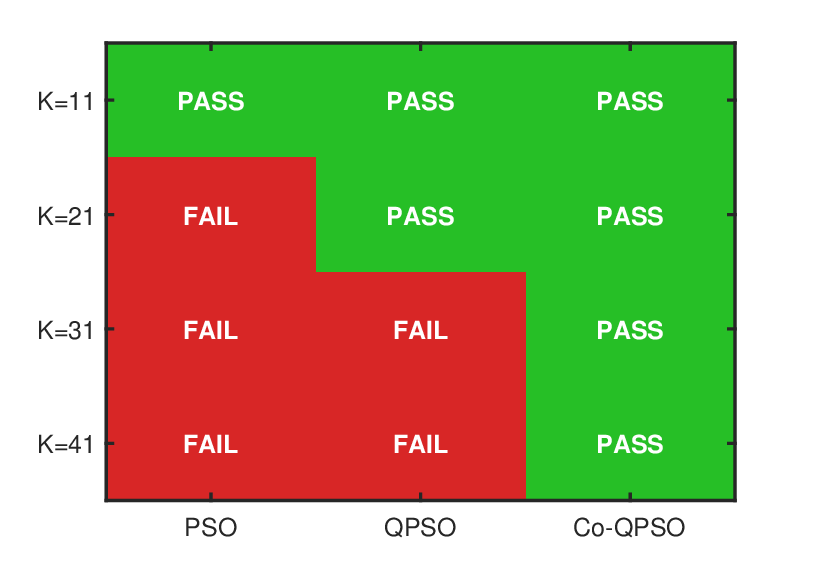}
    \caption{Feasibility map for PSO, QPSO, and Co-QPSO with extended $K$}
    \label{fig:Scalability_2}
\end{figure}

\subsection{Scalability Analysis}
{As shown in Fig. \ref{fig:Scalability_1}, we systematically scaled the total mission duration ($T_{\text{end}}$) from 100s up to 400s, holding the time step constant ($dt = 10$s). This increased the number of time slots $K \in \{11, 21, 31, 41\}$ and expanded the problem dimensionality from 77 to 287 tightly coupled continuous and integer variables. 

The empirical results clearly revealed the superior scalability of the proposed Co-QPSO algorithm. As demonstrated by the feasibility map in Fig. \ref{fig:Scalability_2}, with an increase of problem complexity, classical PSO and standard QPSO suffered from the ``curse of dimensionality,'' becoming trapped in infeasible regions and failing to find viable trajectories for extended missions ($K \ge 21$ and $K \ge 31$, respectively). Conversely, despite the exponentially expanded search space, Co-QPSO consistently converged to feasible, high-quality trajectories at every scale. The total successfully transmitted data scaled proportionately with the mission duration, increasing from $190.06$ Mb ($K=11$) to $625.24$ Mb ($K=41$). This robust performance confirms Co-QPSO's unique viability for planning large-scale, extended UAV missions where traditional standard PSO variants fail.}

\subsection{Convergence Analysis}

Fig.~\ref{fig:conv} shows the convergence behavior of three optimization algorithms: classical PSO, standard QPSO, and the proposed QIEN (Co-QPSO) framework. The results demonstrate the superior convergence and speed of the quantum-inspired approach. Classical PSO failed to locate a feasible solution until late in the search process, remaining trapped in penalty regions for most of the optimization. Similarly, Standard QPSO was found to struggle with the complex constraints, requiring significant iterations to escape infeasibility. By contrast, the proposed Co-QPSO identified valid resource allocation and trajectories faster than the benchmarks, converging to a high-throughput solution by iteration 400. The primary advantage of Co-QPSO lies in the rapid convergence and constraint satisfaction; the cooperative architecture and heterogeneous particle roles enable QIEN to satisfy the stringent error probability constraint $\varepsilon_{\text{tot}}(t_k) \leq 10^{-9}$ hundreds of iterations faster than benchmark methods.

\subsection{Trajectory Optimization}

Figs.~\ref{fig:3d}-\ref{fig:2d} visualize the optimized UAV trajectory in three and two dimensions. The UAV initiated the mission at an altitude near 1055 meters and concludes at similar levels, dynamically modulating its height to balance competing objectives. This elevation intelligently balances the competing objectives of increasing solar harvesting through reduced atmospheric attenuation \eqref{eq:elevation_factor}-\eqref{eq:cloud_attenuation} and maintaining sufficient transfer channel gains \eqref{eq:channel_gain}. The horizontal projection confirms coordinated movement with the ground robot while strictly maintaining the required safety distance from the station.

\subsection{Energy and WPT Performance}
Fig.~\ref{fig:SolarvsWPT} illustrates the fundamental altitude trade-off inherent to solar-powered UAV systems with WPT. The dual-axis shows two opposing physical trends: solar harvesting power increases monotonically with altitude (orange curve) due to reduced atmospheric attenuation ; conversely, WPT received power decays with altitude (green curve) governed by inverse-square path loss. Notably, the actual WPT data points scatter around the theoretical trend, confirming that received power depends not only on altitude but also on the time-varying horizontal distance and dynamic power allocation. This creates a critical design tension: higher altitudes maximize energy sustainability, while lower altitudes enhance WPT efficiency. The distribution of optimized trajectory points demonstrates that the Co-QPSO algorithm dynamically modulates altitude across the 500-2000~m range rather than converging to a static optimum, effectively balancing instantaneous energy availability with link quality. The consistent satisfaction of the robot power requirement (red dashed line) validates that these dynamic altitude choices successfully reconcile harvesting objectives with transfer constraints.

Moreover, Fig.~\ref{fig:energy} confirms successful energy-neutral operation throughout the mission. Solar power harvesting balanced the total power consumption. The algorithm achieves precise energy neutrality, concurrently maintaining a positive margin at each time step, which satisfies constraint (\ref{eq:consumption_margin}) without unnecessary energy waste. The stacked bar chart reveals the power budget breakdown: propulsion is a consistent consumer; communication power $P_1$ to the ground station varies significantly depending on the UAV's link distance; communication power $P_2$ to the robot adjusts based on proximity; and WPT transmission utilizes the remaining budget. The net power balance annotations confirming zero surplus at all instances validate feasibility of perpetual autonomous operation.

Finally, Fig.~\ref{fig:wpt} demonstrates the WPT subsystem performance over time. The received power at the robot successfully satisfied the minimum requirement $P_{\text{robot}} = 2.0$ W while concurrently respecting the maximum limit $P_{\text{WPT}}^{\max} = 2.5$ W throughout the mission. The temporal variation in received power directly correlates with the time-varying UAV-robot distance. Peak WPT delivery occurred when the trajectory geometry was optimal, while the power strictly adhered to the minimum 2.0 W threshold during critical path segments. The Co-QPSO algorithm effectively adjusted the WPT transmit power $P_{\text{tx}}^{\text{WPT}}(t_k)$ to compensate for path loss variations ensuring continuous robot power delivery despite dynamic geometric changes. The dotted red line indicates the robot power requirement was satisfied throughout the mission.
\vspace{-1.5mm}
\section{Conclusion}

In this paper we presented QIEN, a novel framework addressing the joint trajectory and resource optimization problem for solar-powered UAV relay systems supporting autonomous mining robots in post-disaster environments. To this end, we formulated a large-scale MINLP to maximize aggregate data throughput under stringent HRLLC constraints ($\varepsilon_{\text{tot}}(t_k) \leq 10^{-9}$), ensuring energy-neutral UAV operation and continuous WPT to the ground robot. To overcome the severe non-convexity arising from integer blocklength allocation and coupled mobility constraints, QIEN employs a Hybrid Cooperative Co-QPSO algorithm. Using quantum tunneling mechanisms and a heterogeneous swarm architecture, the proposed algorithm effectively balances global exploration and local exploitation. Simulation results demonstrated that QIEN significantly outperforms classical benchmarks, achieving a total throughput of 190 Mb and securing valid resource allocation and trajectories where classical PSO and QPSO failed to find feasible solutions for large number of iterations. The optimized system maintains strict energy-neutral operation with a precise surplus margin of 0 W at every time step and ensures continuous robot power delivery between 2 to 2.2 W, satisfying all mission-critical requirements. Taken together, these findings confirm the practical viability of QIEN for enabling perpetual, autonomous industrial monitoring systems in extreme environments, thus creating a robust foundation for future self-sustaining HRLLC networks.
\vspace{-5mm}
\appendices

\section{Concavo-Convex Analysis}

\begin{theorem}[Non-Convexity of Aggregate Sum Rate]
The time-integrated sum rate objective function is
\begin{equation}
f(\bm{\Phi}) = \Delta t \sum_{k=0}^{K-1} \sum_{i=1}^{2} R_i^*(t_k)
\end{equation}
where \scalebox{0.8}{$\bm{\Phi} = \{x_u(t_k), y_u(t_k), z(t_k), n_1(t_k), P_1(t_k), P_2(t_k), P_{\text{tx}}^{\text{WPT}}(t_k)\}_{k=0}^{K-1}$} represents the joint trajectory, blocklength allocation, and power control variables over $K$ time steps, is non-convex in the optimization variables.
\end{theorem}

\begin{proof}
We establish non-convexity by demonstrating that the Hessian matrix $\nabla^2 f$ is indefinite over the feasible region. The finite blocklength achievable rate for communication link $i \in \{1,2\}$ is given by the following:

\begin{equation}
R_i^* = B\left[ C_i(P_i, d_i) - \log_2(e) \sqrt{\frac{V_i(P_i, d_i)}{n_i}} Q^{-1}(\varepsilon_{\text{target}}) \right],
\end{equation}
where the channel capacity is $C_i = \log_2(1 + \gamma_i)$ with SNR $\gamma_i = P_i h_i/\sigma^2$ and communication channel gain $h_i = \beta_0/d_i^2$, where $\beta_0$ denotes the reference channel gain for information transmission. The channel dispersion is $V_i = 1 - (1+\gamma_i)^{-2}$.

Computing the second-order partial derivative with respect to blocklength $n_i$ yields the following:
\begin{equation}
\frac{\partial^2 R_i^*}{\partial n_i^2} = -\frac{3B \log_2(e)}{4} \sqrt{\frac{V_i}{n_i^5}} Q^{-1}(\varepsilon_{\text{target}}).
\end{equation}
For ultra-reliable regimes with $\varepsilon_{\text{target}} < 0.5$, the term $Q^{-1}(\varepsilon_{\text{target}})$ is positive. While the continuous relaxation is concave ($\frac{\partial^2 R}{\partial n^2} < 0$), variable $n_i$ is constrained to the discrete set $\mathbb{Z}^+$. This disjoint domain breaks the continuity required for convexity, creating a mixed-integer non-linear structure preventing the application of standard convex optimization tools.

Now, the channel gain depends on UAV position through $d_i^2 = (x_u - x_i)^2 + (y_u - y_i)^2 + z^2$, creating bilinear coupling. The mixed partial derivative is as follows:
\begin{equation}
\begin{aligned}
\frac{\partial^2 C_i}{\partial P_i \partial d_i} &= \frac{\partial}{\partial d_i}\left[\frac{1}{\ln(2)} \frac{\beta_0}{\sigma^2 d_i^2 + P_i \beta_0}\right] \\
&= -\frac{2\beta_0 \sigma^2 d_i}{(\sigma^2 d_i^2 + P_i \beta_0)^2 \ln(2)}.
\end{aligned}
\end{equation}
This bilinear term creates saddle points in the $(P_i, d_i)$ subspace. The Hessian matrix restricted to this subspace has eigenvalues of opposite signs, confirming indefiniteness and lack of joint convexity.

Thereafter, the energy sustainability constraint
\begin{equation}
P_{\text{prop}}(z) + P_1 + P_2 + P_{\text{tx}}^{\text{WPT}} + P_{\text{circuit}} \leq A_{\text{solar}} \phi(z) C(z)
\end{equation}
involves the non-linear elevation factor $\phi(z) = \alpha - \beta_s e^{-z/\delta}$. Propulsion term $P_{\text{prop}}$ is convex in $z$, but the harvesting term (right-hand side) is neither convex, nor concave due to the exponential altitude factor. Consequently, the feasible set defined by $P_{\text{cons}}(z) \leq E_{\text{harv}}(z)$ is non-convex. Similarly, the WPT constraint
\begin{equation}
P_{\text{tx}}^{\text{WPT}} \geq \frac{P_{\text{robot}}[(x_u - x_r)^2 + (y_u - y_r)^2 + z^2]}{\eta_{\text{WPT}} \beta_0^{\text{WPT}}}
\end{equation}
defines a region bounded by a convex function. However, the intersection of this with the upper-bound power constraints creates a non-convex feasible region.

Finally, the mobility constraints are
\begin{equation}
\|(x_u(t_{k+1}), y_u(t_{k+1})) - (x_u(t_k), y_u(t_k))\|_2 \leq v_{\max}^{\text{horiz}} \Delta t
\end{equation}
which couple decisions across time steps. While the constraint itself is a convex norm ball, it propagates the non-convexity from power-distance coupling across the entire time horizon $K$. Considering that the objective function has an indefinite Hessian and the constraints involve non-convex functions and integer variables, Problem P1 is fundamentally non-convex.
\end{proof}
{\section{Algorithm Justification}
\subsection{Theoretical Justification for Algorithm Selection}
\label{sec:theoretical_justification}

While there are various optimization methodologies for UAV trajectory planning, the specific architectural constraints of the QIEN framework require a heavy-tailed metaheuristic approach of Co-QPSO. In what follow, we theoretically justify this selection against both gradient-based and classical evolutionary alternatives.

\subsubsection{Limitations of Convex Approximations (SCA/BCD)}
Problem P1 is formulated as an MINLP featuring the integer blocklength allocation vector $\bm{n}(t_k)$. Classical approaches using SCA or Block Coordinate Descent (BCD) require continuous relaxation of these integer constraints. However, while continuous relaxation is mathematically sound in asymptotic Shannon-capacity models, it fails in the Finite Blocklength (FBL) regime. For HRLLC, $n_i(t_k)$ is inherently small. Rounding a relaxed continuous solution $\tilde{n}_i(t_k)$ to the nearest integer generates substantial quantization errors. Because of the high sensitivity of the inverse Q-function $Q^{-1}(\cdot)$ in Eq. \eqref{eq:achievable_rate}, this rounding frequently violates the stringent $10^{-9}$ reliability constraint \eqref{eq:c6}, thus rendering the SCA-derived trajectory physically infeasible. By contrast, Co-QPSO processes $\bm{n}(t_k)$ as discrete variables natively, thus entirely bypassing relaxation penalties.

\subsubsection{Structural Incompatibility with DRL and Lyapunov Optimization}
Alternative frameworks such as Deep Reinforcement Learning (DRL) and Lyapunov optimization present structural mismatches when applied to the QIEN system. DRL typically optimizes a policy $\pi_{\theta}$ by adopting the following policy gradient:
\begin{equation}
\nabla_{\theta} J(\theta) = \mathbb{E}_{\pi_{\theta}}[\nabla_{\theta} \log \pi_{\theta}(a|s) Q^{\pi}(s,a)],
\label{eq:policy_gradient}
\end{equation}
where $J(\theta)$ is the expected return, $s$ and $a$ are the state and action vectors, and $Q^{\pi}(s,a)$ is the action-value function representing the long-term cumulative reward. While $Q^{\pi}(s,a)$ is frequently estimated using the recursive Bellman optimality equation, the resulting gradient updates function as local search mechanisms. In the presence of the fragmented feasible manifold created by altitude-dependent solar harvesting and distance-dependent WPT, such updates tend to converge to sub-optimal local attractors. Furthermore, DRL handles constraints via soft penalties in the reward function, which cannot strictly guarantee the $10^{-9}$ reliability threshold that is essential for safety-critical mining surveillance. Similarly, Lyapunov optimization is designed for long-term average stability by minimizing a drift-plus-penalty expression:
\begin{equation}
\Delta(L(t)) + V \mathbb{E}[f(t)|L(t)],
\label{eq:lyapunov_drift}
\end{equation}
where $L(t)$ is the Lyapunov function (frequently representing energy queues), $\Delta(L(t))$ is the drift, $V$ is a control parameter, and $f(t)$ is the penalty function representing the negative sum-rate objective. Since this methodology is fundamentally myopic as it makes decisions on a per-slot basis to maintain queue stability, it fails to account for the global temporal coupling of  mobility constraints in Eq. \eqref{eq:c12}-\eqref{eq:c13} over the finite mission duration $T_{\text{end}}$. By contrast, Co-QPSO treats the entire $K$-step trajectory as a single high-dimensional decision vector, ensuring that the solar energy harvested in early time steps is strategically preserved for high-power WPT requirements in later intervals.

\subsubsection{Probabilistic Superiority over GA and DE}
Classical metaheuristics, such as Genetic Algorithms (GA) and Differential Evolution (DE), struggle with the discontinuous feasible manifolds generated by the intersection of strict WPT (37i) and energy sustainability (37h) constraints. To bridge disconnected feasible regions (the ``invisible wall'' phenomenon), an algorithm requires high-magnitude spatial jumps. 

In standard GA featuring Gaussian mutation, the probability density function for a spatial jump of distance $x$ is proportional to the normal distribution. The probability of making a large jump to escape a local optimum decays quadratically as shown below:
\begin{equation}
P_{\text{GA}}(|X| > x) \sim \int_{x}^{\infty} e^{-t^2} dt.
\end{equation}
In DE, mutation is dictated by the difference vectors between existing population members. Since the population begins to cluster around a local optimum, the variance of the difference vectors approaches zero, which mathematically eliminates the capacity for large-scale exploration.

Conversely, the position update mechanism in QPSO is driven by the stochastic term $\ln(1/u)$, where $u \sim U(0,1)$. By the inverse transform sampling theorem, the negative logarithm of a standard uniform variable yields a standard exponential distribution. Factoring in the symmetric random sign multiplier $s \in \{-1, 1\}$, the step size mathematically follows a double-exponential Laplace distribution. The tail probability for QPSO decays only linearly in the exponent (see Eq. \ref{eq:qpso_bound}):
\begin{equation}\label{eq:qpso_bound}
P_{\text{QPSO}}(|X| > x) \sim \int_{x}^{\infty} e^{-t} dt = e^{-x}.
\end{equation}
Since $e^{-x} \gg e^{-x^2}$ for large values of $x$, the Laplace distribution inherently possesses heavier tails. Therefore, according to global convergence theory for stochastic search algorithms, QPSO maintains a strictly higher mathematical probability of tunneling through expansive infeasible regions, as compared to GA and DE, thus structurally preventing premature convergence in highly constrained MINLPs.}

\bibliographystyle{IEEEtran}
\bibliography{Bibliography-1.bib}

\end{document}